\documentclass[preprint,12pt]{elsarticle}

\usepackage{hyperref}
\usepackage{amssymb}
\usepackage{amsthm}
\usepackage{amstext}
\usepackage{amsmath}
\usepackage{epsfig}
\usepackage{epstopdf}

\biboptions{comma,sort}

\journal{Astroparticle Physics}

\begin{document}
\DeclareGraphicsExtensions{.pdf,.gif,.jpg}

\begin{frontmatter}

\title{First detection of Extensive Air Showers by the TREND self-triggering radio experiment}

\author[subatech]{D. Ardouin}
\author[lpc]{C. C\^arloganu}
\author[subatech]{D. Charrier}
\author[ihep]{Q. Gou}
\author[ihep]{H. Hu}
\author[gucas]{L. Kai}
\author[subatech]{P. Lautridou}
\author[ihep,naoc,lpnhe]{O. Martineau-Huynh\corref{cor}}
\ead{omartino@in2p3.fr, }
\author[lpc]{V. Niess\corref{cor}}
\ead{niess@in2p3.fr}
\author[subatech]{O. Ravel}
\author[naoc]{T. Saugrin\corref{cor}}
\ead{thomas.saugrin@gmail.com}
\author[naoc]{X. Wu}
\author[ihep]{J. Zhang}
\author[ihep]{Y. Zhang}
\author[naoc]{M. Zhao}
\author[gucas]{Y. Zheng}

\address[subatech]{SUBATECH, Ecole des Mines, CNRS/IN2P3 and Universit\'e de Nantes, 44307 Nantes, France}
\address[lpc]{Clermont Universit\'e, Universit\'e Blaise Pascal,CNRS/IN2P3, Laboratoire de Physique Corpusculaire, BP 10448, F-63000 Clermond-Ferrand,vFrance}
\address[ihep]{Key Laboratory of Particle Astrophysics, Institute of High Energy Physics, Chinese Academy of Sciences, Beijing 100049, P.R. China}
\address[gucas]{Graduate University of Chinese Academy of Science, Beijing 100049, P.R. China}
\address[naoc]{National Astronomical Observatories of China, Chinese Academy of Science, Beijing 100012, P.R. China}
\address[lpnhe]{Laboratoire de Physique Nucl\'eaire et des Hautes Energies, CNRS/IN2P3 and Universit\'e Pierre et Marie Curie, 75252 Paris Cedex, France}
\cortext[cor]{Corresponding authors.}

\begin{abstract}
An antenna array devoted to the autonomous radio-detection of high energy cosmic rays is being deployed on the site of the 21~cm array radio telescope in XinJiang, China. Thanks in particular to the very good electromagnetic environment of this remote experimental site, self-triggering on extensive air showers induced by cosmic rays has been achieved with a small scale prototype of the foreseen antenna array. We give here a detailed description of the detector and present the first detection of extensive air showers with this prototype.
\end{abstract}


\end{frontmatter}

\section{ Introduction }

The Tianshan Radio Experiment for Neutrino Detection (TREND) is a Sino-French collaboration proposing to build a large radio array in order to search for Ultra High Energy (UHE) neutrinos. These neutrinos may undergo charged current interactions with the matter below the Earth surface. In order to detect these particles, it has been proposed~\cite{Zas:1992, Fargion:1999, Fargion:2001, Vannucci:2001, Bertou:2002, Letessier:2000,Zas:2005} to search for very inclined or Earth-skimming extended air showers (EAS) induced by the decay of tau leptons produced by charged current neutrino interactions in the Earth. The motivation and status of the radio-detection of ultra-high energy cosmic rays (UHECRs) and UHE neutrinos will be briefly discussed in \autoref{sec:status}. We will state in \autoref{sec:proto} the very good electromagnetic conditions at the TREND site and present the proof of principle that the TREND antenna array can detect EAS in a stand-alone mode. This result constitutes a first and important milestone for the purpose of UHE neutrino detection. Given the results and performances obtained with the TREND radio-detector, the motivations for a UHE neutrino search with TREND are evoked in \autoref{sec:outlook}.

\section{ Development of the radio technique in the context of  UHECR and neutrino detection }
\label{sec:status}
\subsection{ Radio antennas for the detection of UHECRs}

Despite intense experimental effort over the last twenty years, the nature and origin of cosmic rays above the knee ($\sim3\times10^{15}$~eV) are still uncertain. One of the major experimental challenges is the very low flux of high energy cosmic rays, thus requiring the use of very large detection surfaces or volumes in order to acquire statistically significant results~\cite{Bluemer:2009}. Very encouraging results recently obtained by the Pierre Auger Observatory~\cite{Abraham:2009a, Abraham:2009b} indicate the relevance of hybrid detection systems deployed over extremely large surfaces for the study of UHECRs spectrum, composition and origin. \\
\indent Parallel to this, recent works by the LOPES and CODALEMA experiments have produced interesting results for the characterization of cosmic rays above 10$^{16}$~eV, using radio antenna arrays triggered by ground detectors of EAS~\cite{Falcke:2005, Ardouin:2005, Ardouin:2006, Ardouin:2009a, Horneffer:2006, Apel:2009}. These results indicate that radio detection could become an alternative or complementary technique to the systems used presently for the detection of UHECRs. One of the attractive aspects of radio detection is the very low cost of antennas and its easiness of deployment over large areas. However, the potential of the radio-detection technique can be fully exploited only if working as a self-triggering system. We present in this paper the proof of principle for such a self-triggering mode for a radio detector of EAS, validated on a small-scale prototype of the TREND setup.

\subsection{ Hunting UHE neutrinos with radio antennas }
Radio emission by particle showers was first predicted by Askaryan in the early 60's~\cite{Askaryan}. The use of the radio technique for the search of neutrino-induced showers in a natural dense medium was proposed~\cite{Zheleznykh,Gusev} in the early 80's, following Askaryan's suggestions. Latter in the 90's, several projects set up antenna arrays to search for radio waves from neutrino-induced showers in natural dense medium such as ice, rock salt formations or lunar regolith (see for instance~\cite{Frichter, Gorham:2004, Barwick, Gorham:2002}). Even if some very exciting results have recently been released (see for instance~\cite{anita:2010}), only limits on the UHE neutrino fluxes have been set so far by these experiments. Though the attenuation length of radio waves in solids are large enough (typically a few hundred meters) to equip large detection volumes, the achievement of a better sensitivity requires an increase in detection surface and number of (deep)-embedded detectors, which could be expensive and time consuming, in the case for example of Antarctic ice experiments. \\
\indent The setup of radio antenna arrays at the surface of the Earth looking for very inclined EAS may represent for UHE neutrino detection a cost-effective alternative both to these projects and to the well-advanced EAS ground and fluorescence detection techniques also used for this search~\cite{Abraham:2009c, Gora:2009}. Thanks to the important breakthroughs~\cite{Falcke:2005, Ardouin:2005} performed in air-shower radio detection, the idea of setting  antennas detecting radio emission in the atmosphere has triggered intense interest lately~\cite{Brusova:2007, Ardouin:2009b,Carloganu:2009}. Yet, no complete evaluation of this technique's potential has been performed so far. One of the main challenge in identifying nearly horizontal showers lies in the existence of anthropic noise sources close to the horizon, together with a possible influence of distant storms on the atmosphere electromagnetic condition. A detailed analysis of the sensitivity of the TREND antennas to nearly horizontal showers is beyond the scope of this article. Nevertheless, we will state in this paper the uniqueness of the electromagnetic environment of the TREND site, remotely located and surrounded by very close and high (up to 5000 m a.s.l.) mountains.
%

\section{The TREND detection setup and analysis procedure}
\label{sec:proto}
The TREND project uses a large part of the existing infrastructures of the 21~cm array (21CMA) detector.

\subsection{The 21CMA interferometer}
\label{21cma}
\begin{figure}
\begin{center}
\includegraphics[width=9.5cm,height=6.5cm]{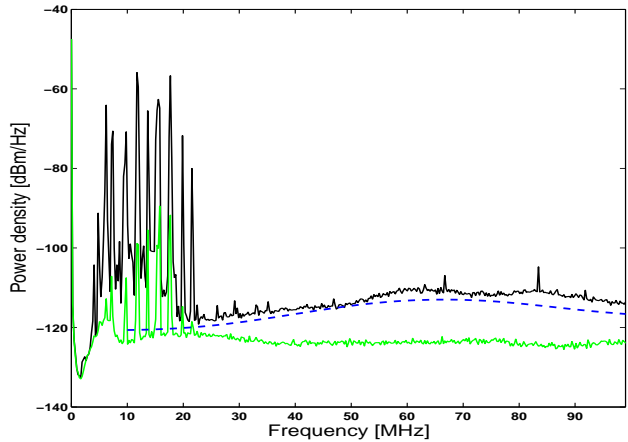}
\caption{ \label{fig:ulastai-noise-spectrum} \textit{
  Noise power spectrum measured with a 21CMA antenna and a 48~dB low noise amplification (in black). The green curve shows the spectrum recorded when the input cable is disconnected from the antenna. The dashed blue curve is the simulated response of an ideal system to the galactic signal only. The frequency range of this measurement is 0-100~MHz. Few contributions beyond the galactic emission are found above 25~MHz. }}
\end{center}
\end{figure}

\begin{figure}
\includegraphics[width=6.3cm,height=6.3cm]{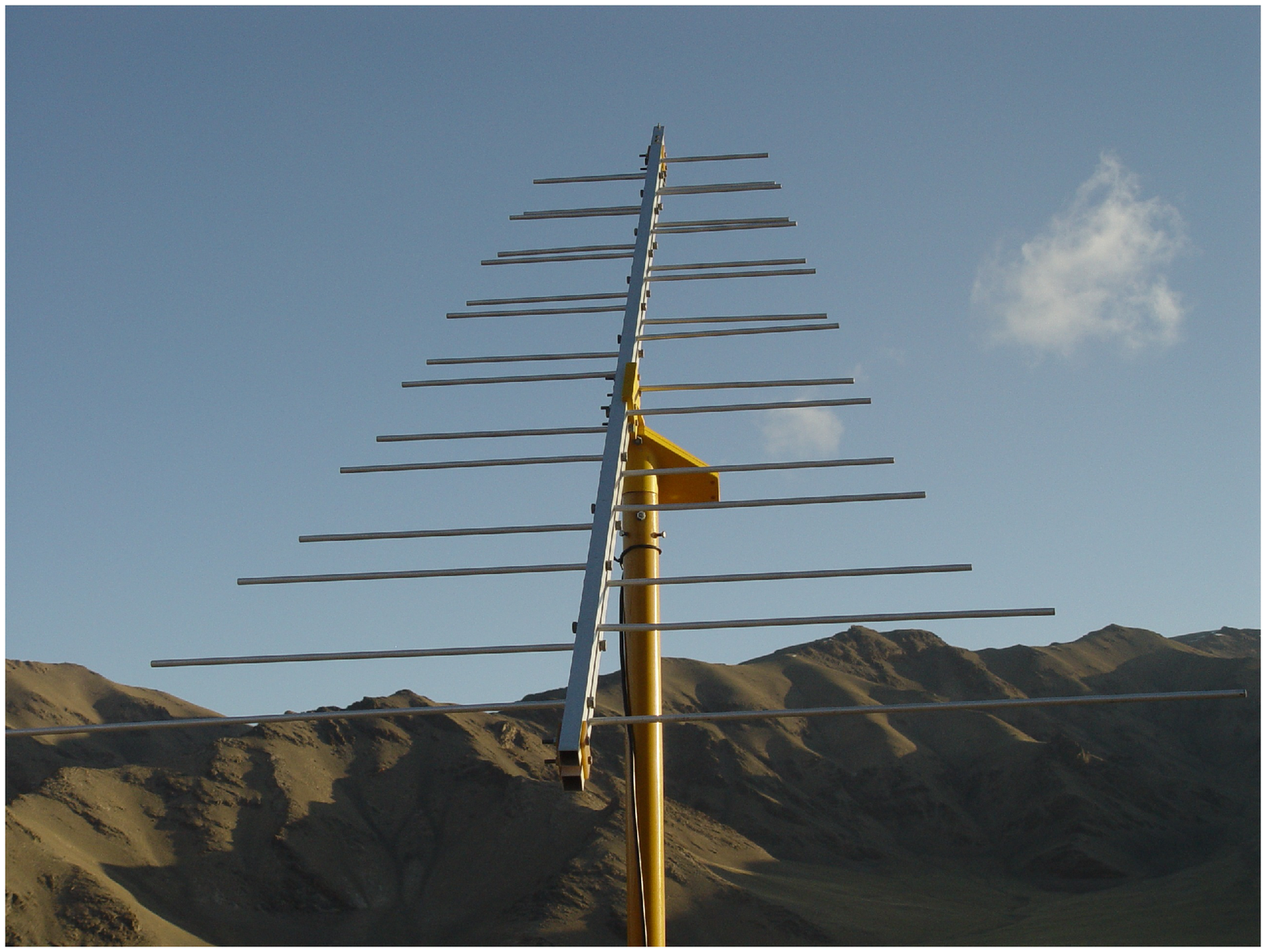}
\includegraphics[width=8cm,height=6.3cm]{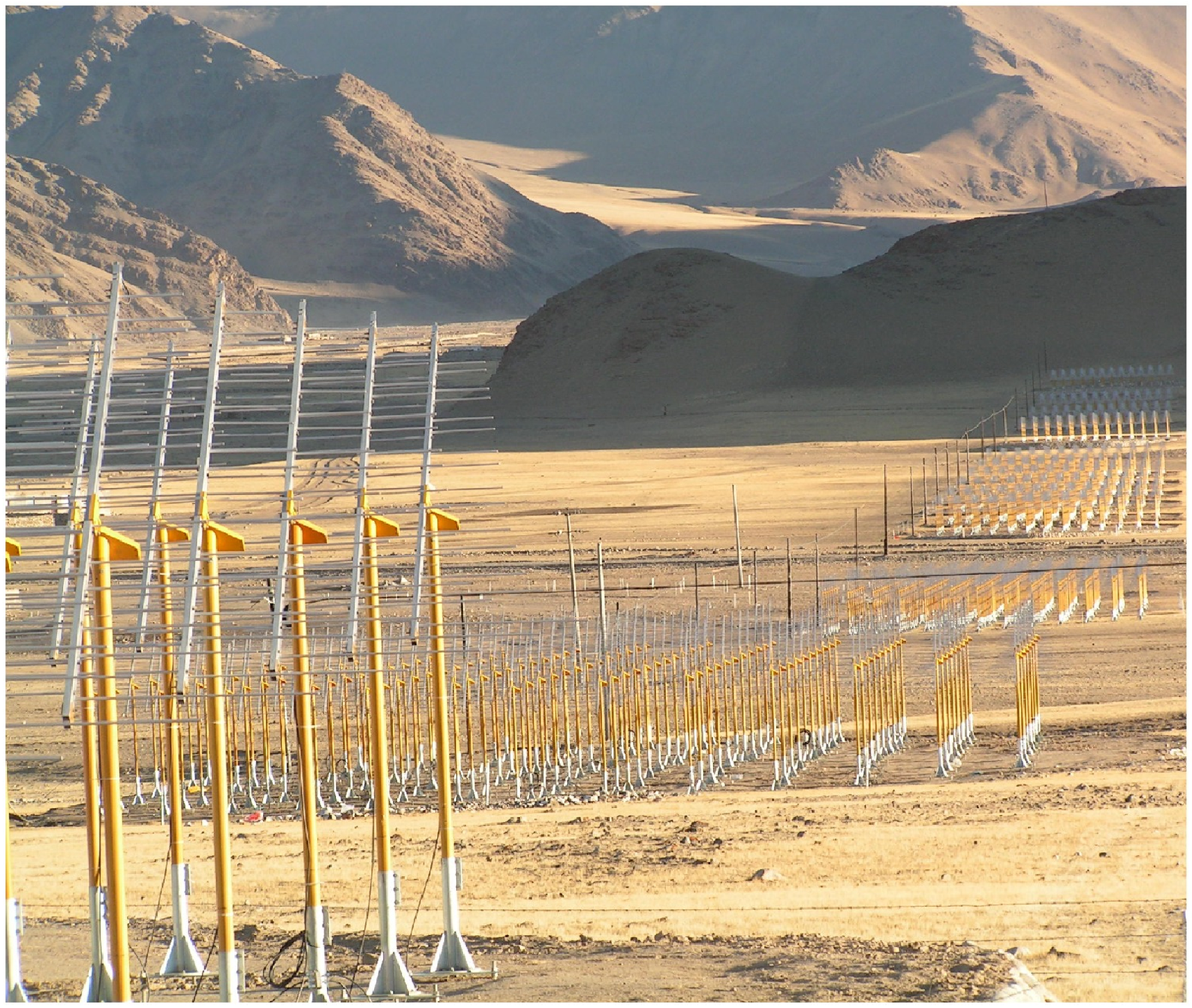}

\caption{ \label{fig:antenna} \textit{
  Left~: picture of a 21CMA log-periodical antenna used in the TREND prototype setup. Right: picture of several log-periodical antennas and pods from the South-North arm. The size of a pod is 20~m$\times$32~m. }}
\end{figure}

\begin{figure}
\begin{center}
\includegraphics[width=8.5cm,height=8cm]{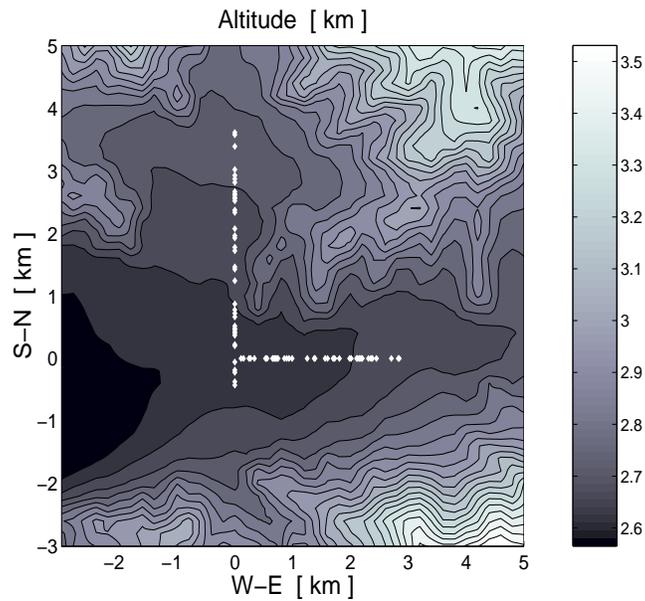}
\caption{ \label{fig:21cma-layout} \textit{
  The 21CMA detector layout. Pods positions are indicated with white diamonds. The color code gives the altitude a.s.l. in kilometers. The x and y axis correspond respectively to the West-East and South-North directions. The cross-point of the two baselines is taken as the origin of the referential. }}
\end{center}
\end{figure}

The 21~cm array is a radio-interferometer aiming at the study of the epoch of reionization \cite{Furlanetto:2006}. It was completed in 2007 by the National Astronomical Observatories of China in the Tianshan mountain range (XinJiang, China)~\cite{21CMA:url}, at an altitude of 2650~m.\\
\indent Due to its remote location, the 21CMA site benefits from a very clean electromagnetic environment, nearly free of stable sources with significant emission beyond the galactic background noise. Only two close-by 10~kV power lines and a nearby railway are localized background sources. The quality of the electromagnetic environment is illustrated in \autoref{fig:ulastai-noise-spectrum}, which represents the radio background observed with a 21CMA antenna directly plugged into a 48~dB low-noise amplification system. This spectrum was recorded at the galactic signal maximum (18h00 local sideral time). For the purpose of illustration, the simulated response of this system to the galactic emission is also plotted in \autoref{fig:ulastai-noise-spectrum}. \\
%
\indent The 21CMA is composed of 10160 antennas, grouped in 80 pods of 127 antennas each. Commercial log-periodical antennas have been used. They consist of 18 side-by-side parallel dipoles (see \autoref{fig:antenna}) with geometrical patterns designed to work in the 50-200~MHz bandwidth -which corresponds to the range of the 21~cm hydrogen emission line for the epoch of reionization~\cite{Furlanetto:2006}- and achieving antenna gains of $\sim$10~dBi. The dipoles are positioned horizontally, resulting in a polarization along the East-West direction, as the antennas are orientated towards North. An inclination of 47$^\circ$ with respect to ground (so that antennas point towards the Polar Star) corresponds to a maximum of the gain pattern for $\sim$40-60$^\circ$ zenith angle over the 50-200~MHz frequency band. Over this frequency range, the zenithal beamwidth at 3~dB varies between 20 and 60$^{\circ}$, while the other antenna characteristics are those of standard log-periodical antennas.  These antennas do not have the required sensitivity towards the horizon  for the foreseen search for nearly horizontal showers induced by $\tau$ leptons (see \autoref{sec:outlook}). They were nevertheless successfully used for the present stage of the TREND experiment, which consists in the validation of the autonomous triggering mode of the radio array on EAS induced by cosmic rays.  \\
%
\indent The 21CMA pods form two perpendicular baselines orientated along the North-South and East-West directions.  They lie in the bottom of two high-altitude valleys and extend for 4.0 and 2.8~km respectively (see \autoref{fig:21cma-layout}). For each pod, the sum of the analog signals from the 127 antennas is continuously fed into an optical transmitter placed in the middle of the pod. Each transmitter is connected through an optical fiber to an acquisition room situated in the center of the East-West arm. The signals from the 80 fibers are then digitized in parallel by 8-bits ADCs at a sampling rate of 200~MSamples/s. The data of each pod are finally buffered on 200~MB disks, where the signal processing is performed.
%

\subsection{The TREND prototype setup}
\label{sec:daq-description}
The driving concept of the TREND experiment is that the 21CMA can be used as a self-triggered detector for EAS through minor adaptations of the existing setup. These adaptations mainly consist in using a single antenna instead of a group of 127 as the unit detector, and setting up an on-line software program triggering the writing of the data to the disk independently for each channel.  \\
\indent After preliminary tests carried out in 2008, a prototype composed of six 21CMA-type log-periodical antennas was deployed in January 2009 in order to validate this concept. It was tested for several months with various antenna layouts. The antennas were oriented towards North, thus polarized along the East-West axis. Each antenna signal is fed into an optical transmitter placed in the middle of the associated pod after a 64~dB low-noise amplification and a 50-100~MHz filtering. As for 21CMA, the signals are sampled by 200 MSamples/s ADCs running in parallel in the acquisition room after being transferred through optical fibers. It should be pointed out that, as two optical fibers are connecting a pod to the acquisition room, TREND can run in parallel with 21CMA without any interference between the two experiments. \\
\indent Following the transient identification method presented in \cite{Ardouin:2005}, a software trigger is set for each digitized signal, and a waveform composed of 2048 samples (10.24 $\mu$s) is written to disk whenever a sample value exceeds a threshold set as a multiple $N$ of $\sigma$, the instantaneous noise level of the antenna. $\sigma$ is calculated every second over 2048 consecutive samples of the buffered data.  The multiplicative coefficient $N$ is chosen at the beginning of the data acquisition such that the antenna triggers at a rate around 1~Hz. A typical value for $N$ is 6.5, for which the expected trigger rate is equal to 0.2~Hz for a sampling frequency of 200~MSamples/s, assuming a stationary Gaussian noise distribution. During quiet periods, we indeed measured trigger rates below 1~Hz for $N$ = 6.5.

\subsection{Antenna sensitivity}
\label{sec:antenna-sensitivity}
\begin{figure}
\includegraphics[width=7cm,height=5cm]{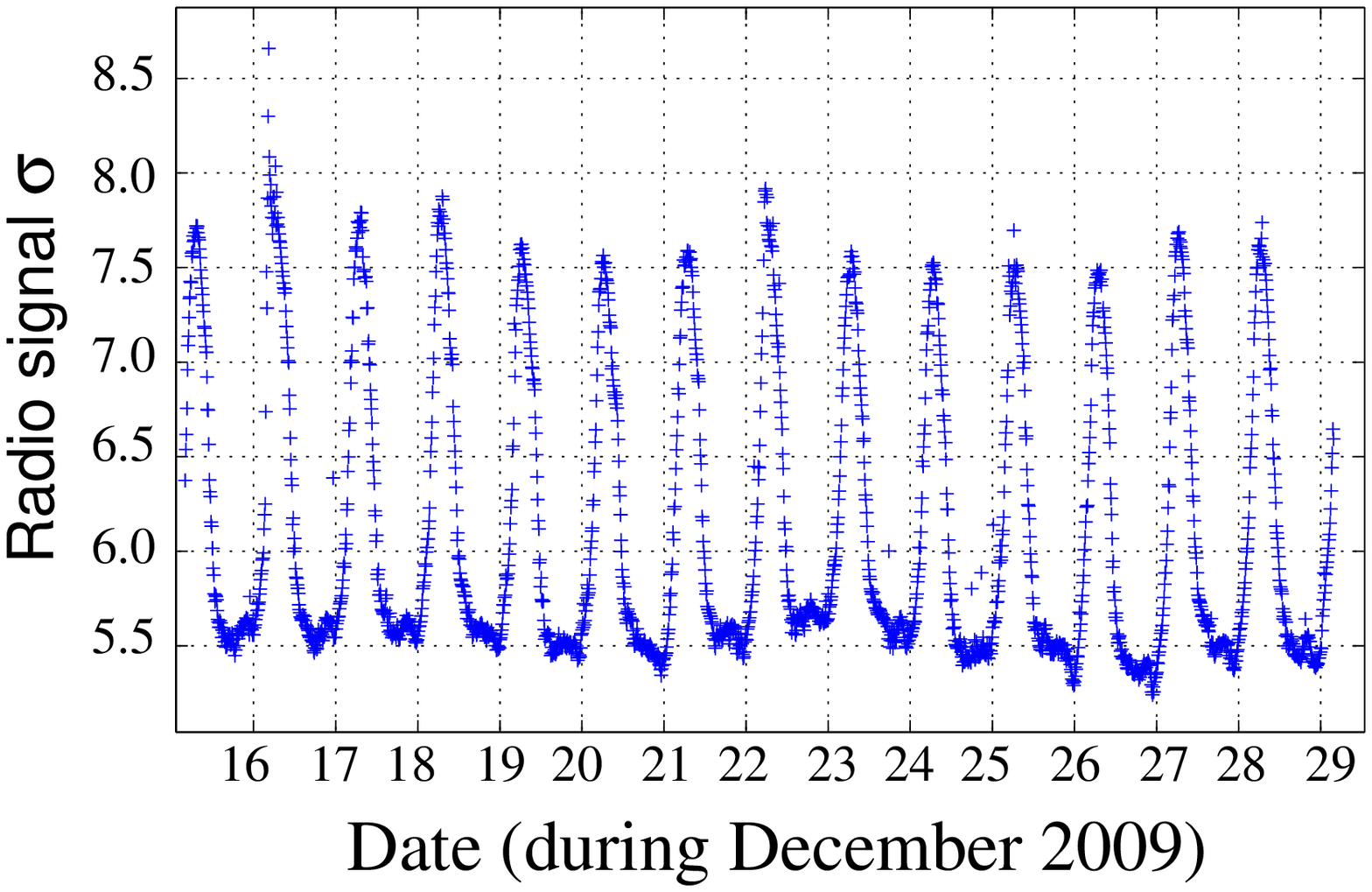}
\includegraphics[width=7cm,height=5cm]{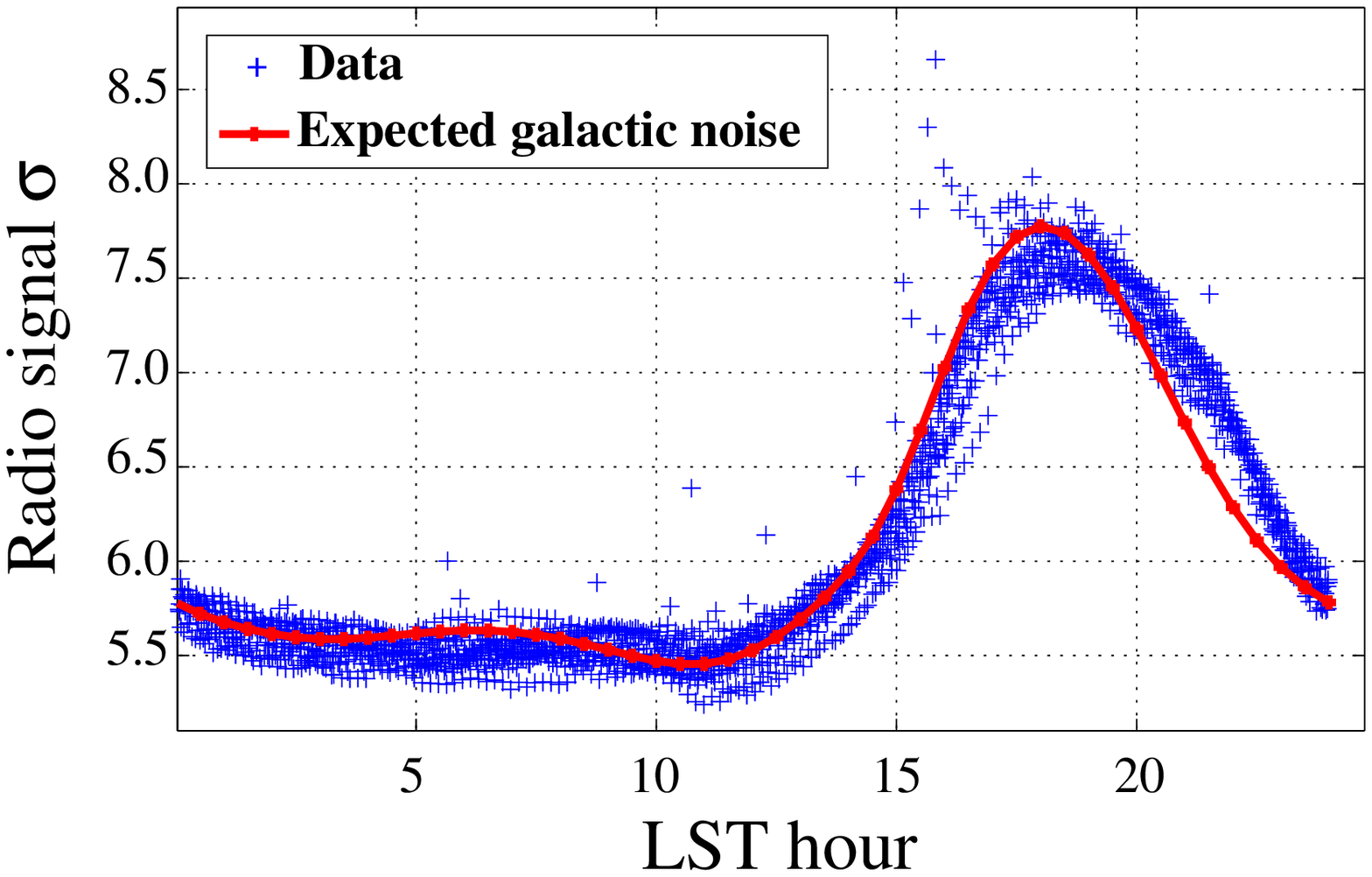}
\caption{ \label{fig:galactic-noise} \textit{
  Left: evolution of the noise level (in units of least significant bit) for one TREND  antenna over a period of 14 days. Each point corresponds to the signal standard deviation measured over a period of 5~ms in the frequency range 50-100~MHz. Right: same data, but with time expressed as local sideral time. The period of high noise level corresponds to the crossing of the visible sky by the Galactic plane. Superimposed is the expected signal for an antenna pointing to the North, simulated as done in \cite{Lamblin:2008}. The slight mismatch in time is explained by a modest tilt of the antenna from full North towards West. }}
\end{figure}

A 5~ms subset of data is recorded every 5 minutes on all antennas in order to monitor the system. The evolution of the noise level of the antennas in particular gives a good diagnostic of  their sensitivity. The CODALEMA experiment has indeed shown that the noise level on a radio antenna should follow a periodical evolution, with a higher value when the galactic plane is visible in the sky, as the Milky Way is the dominant emitter in the sky at radio frequencies \cite{Lamblin:2008}. \\
\indent This feature is clearly observed on all TREND prototype antennas, without any correction nor time averaging (see \autoref{fig:galactic-noise}), proving the outstanding quality of the 21CMA radio environment. Moreover, having identified a common physical source shining uniformly over the whole array allows in principle to calibrate the antennas responses. This method has not been used so far, and the amplitudes of the pulses are presently normalized by applying to their raw value a calibration factor equal to 1/$\sigma$, $\sigma$ being here again the instantaneous antenna noise level measured at the end of the acquisition chain.  As the electronic chain is not the dominant noise source, $\sigma$ is a reliable measure of the antenna sensitivity to electromagnetic signals. This method provides a quick way of performing a relative calibration of the signals amplitudes.
%

\subsection{Selection of coincident antenna triggers}
\label{sec:coincidences-selection}
The first step of the off-line data analysis consists in the search for coincident triggers between several antennas. This is done by ordering in time all triggers from a run, and selecting in this list all consecutive values which may be associated with the same electromagnetic source. Trigger times for antennas $i$ and $j$ are causally linked if they follow the condition:
%
\begin{equation}
\label{eq:causality-criterium}
|t_i-t_j|\leq \frac{d_{ij}}{c} \times T,
\end{equation}
$d_{ij}$ being the distance between these two antennas and $c$ the velocity of light. Trigger times $t_i$ and $t_j$ are defined here as the times when the 2048-sample waveforms reach their absolute maximum on antennas $i$ and $j$ respectively. The times are corrected for delays associated with the signal propagation through the optical fibers and the cables. $T$ is a factor introduced to allow for possible discrepancies between  $t_i$ and the 'true' trigger time (that is, the actual time at which the electromagnetic wave touches antenna $i$). A value $T$=1.1 is chosen in this analysis, which is a safe factor considering the TREND timing resolution (see \autoref{sec:reconstruction-performances}). \\
%

\begin{figure}
\includegraphics[width=6cm,height=5.5cm]{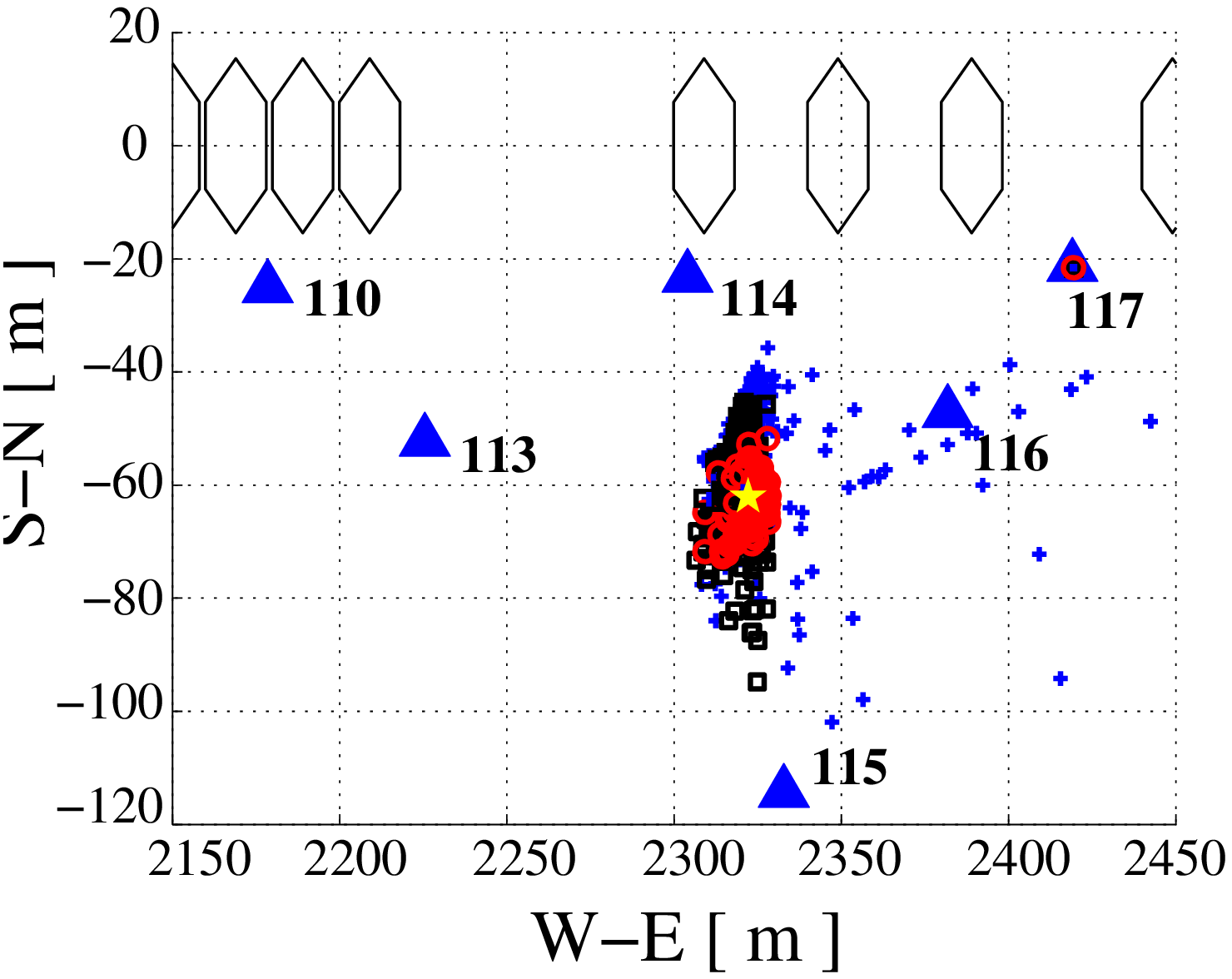}
\includegraphics[width=6cm,height=5.5cm]{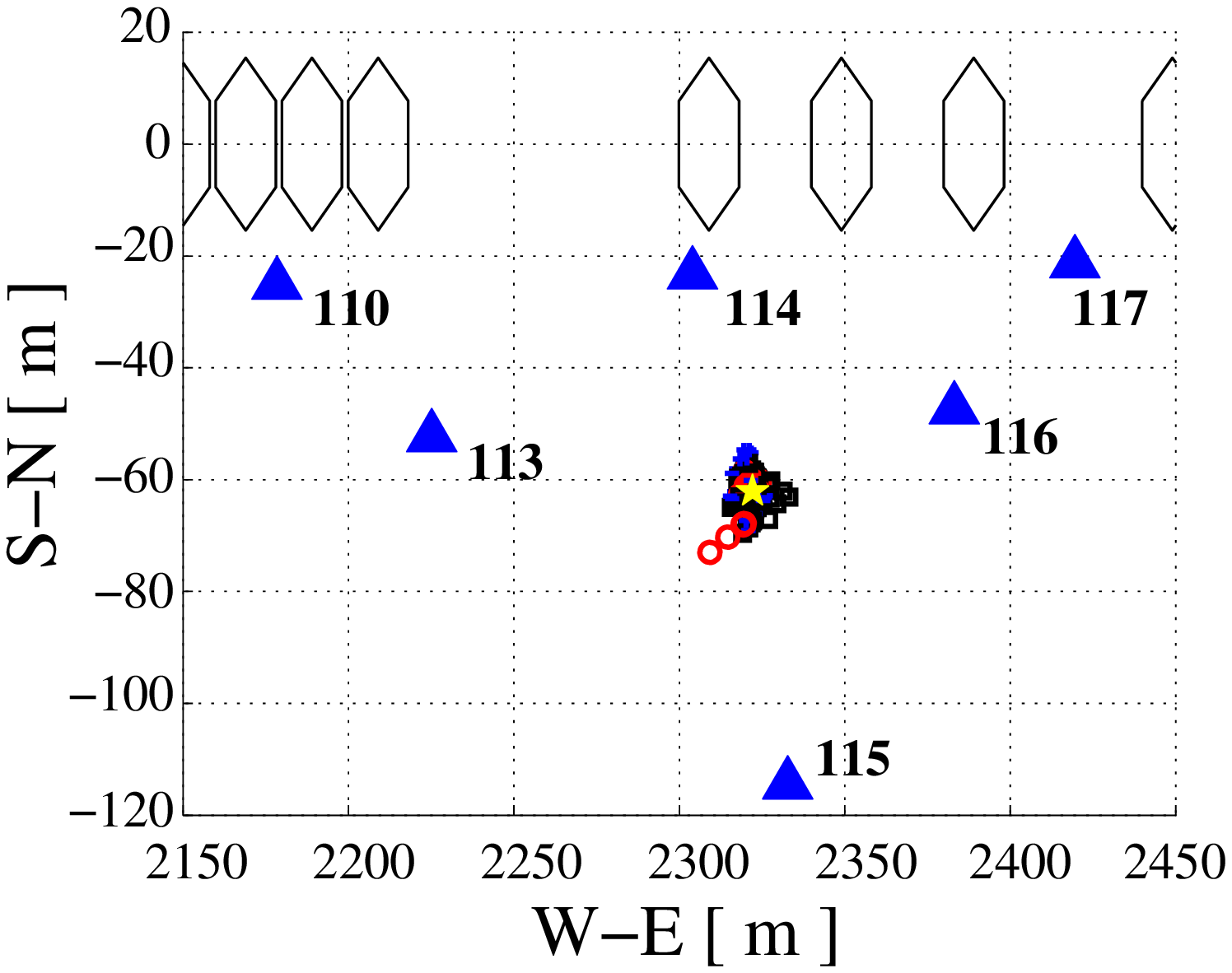}
\caption{ \label{fig:car-calibration} \textit{
  Left:  one of the ground layouts used for the 6-antennas TREND prototype represented in the cartesian referential defined in \autoref{fig:21cma-layout}. The positions of the antennas are indicated by blue triangles, and are labeled by their pod number. The positions of the pods are indicated by hexagons. Also plotted is the reconstructed position of the source for 1642 events recorded during a static source calibration run (see \autoref{sec:reconstruction-performances}). They are shown as blue crosses, black squares and red circles for events triggering  4, 5 and 6 antennas respectively. The true source position is indicated with a yellow star. Right:  reconstructed positions of the source for the same data, after the cross-correlation treatment described in \autoref{sec:coincidences-selection} has been applied. }}
\end{figure}
\indent If at least four consecutive triggers from four different antennas follow the criterion defined in \autoref{eq:causality-criterium}, we consider that these triggers are in coincidence and form an event of multiplicity $L$. We then apply to the corresponding waveforms a cross-correlation treatment. Its principle is to determine, for the $L$($L$-1)/2 pairs of signals $S_i$ and $S_j$ present in the event, the delay $\tau_{ij}$ which maximizes the cross-correlation coefficient $\Gamma_{ij}$:
%
\begin{equation}
\Gamma_{ij}(\tau_{ij})=\frac{1}{\sqrt{\int{{S_i}^2}\times\int{{S_j}^2}}}\int{S_i(t)}S_j(t-\tau_{ij})dt,
\end{equation}
where $S_i$ and $S_j$ actually correspond to a subset of 200 samples around $t_i$ and $t_j$ respectively. For signals associated with close-by sources (for example a train moving on the railroad, or a car standing in the middle of the array with its engine running as in \autoref{fig:car-calibration}), it appeared that this treatment was giving the best results when the signal envelopes were used rather than the raw waveforms. In practice however, treatment speed optimization led us to perform the cross-correlation treatment on the absolute values of the signals rather than the envelopes, the performances of the former being only slightly worse. By solving the system of $L$($L$-1)/2 equations $t_i^*-t_j^* = \tau_{ij}$ according to the least squares method, we finally obtain all corrected values $t_i^*$ for antenna trigger times relative to an arbitrary reference $t_{i0}^*$=0. \\
\indent It was observed that for sources standing at fixed positions or moving along well-defined trajectories, this treatment significantly improves the precision of the reconstructed source position, compared to using the raw trigger times (see \autoref{fig:car-calibration}). An explanation could be that the $L$ trigger times are deduced from an over-constrained system of $L$($L$-1)/2 equations. It can also be pointed out that the complete waveform information is taken into account in the cross-correlation treatment, while only one sample (the one with maximum amplitude) is used in the determination of the raw trigger time. As noise and signal have comparable amplitudes, it is not surprising that the later method is less precise. Note nevertheless that the difference between the two treatments is smaller than 3 samples (15~ns) for most signals. Given its excellent results, the cross-correlation treatment was included in the data analysis chain. It should be kept in mind that the waveforms associated with these sources are usually characterized by rather long time extensions (several hundreds of nanoseconds). The effect of the cross-correlation treatment may therefore be not as significant in the case of prompt signals such as the ones expected for the electromagnetic emissions associated with EAS.

\subsection{Reconstruction algorithms}
\label{sec:reconstruction}
The following step in the analysis chain consists in the reconstruction of the direction of origin of the events formed by coincident triggers. \\
\indent Let us first consider the case of two antennas triggering on a wave propagating spherically from a point source. Their trigger times difference is noted $\Delta t_{sph}$. If we perform the reconstruction of the signal direction of origin assuming a plane wave hypothesis, the resulting trigger time difference $\Delta t_{plan}$ will be wrong by a quantity $\epsilon = \Delta t_{plan}- \Delta t_{sph}$ given at leading order in $d/R$ by~:

\begin{equation}
\epsilon = \frac{d^3}{8cR^2}\cos \theta \sin^2 \theta,
\end{equation}
where $c$ is the velocity of light, $R$ the distance to the source and $\theta$ the angle between the source direction and $\vec{d}$, the vector joining the two antennas. Taking $d$ = 250~m (the maximum extension of the TREND prototype array, see \autoref{fig:car-calibration}), we find that $\epsilon<$10~ns (our estimated experimental resolution on the trigger time measurement, see \autoref{sec:reconstruction-performances}) for R$\geq$500~m. According to this calculation, the discrimination between spherical and plane wavefronts is hardly possible for sources further than 500~m with the TREND setup, because of its limited extension and modest timing resolution. Monte-Carlo studies, carried out with the complete setup geometry, yield similar results. Note however that with larger setups, the distance up to which wavefronts curvature radii can be reconstructed will increase. \\
\indent The maximum development point of a vertical $10^{18}$~eV EAS is roughly situated at an altitude of 3200~m asl, 500~m above ground at the TREND location. The curvature radius of an EAS radio wavefront can therefore be reconstructed only for non-inclined EAS with energies above $10^{18}$~eV.
 This most certainly corresponds to a very small fraction of the showers detectable by this setup. In the vast majority of cases, the direction of propagation of an EAS can therefore safely be reconstructed under the plane wave hypothesis with the TREND prototype. This reconstruction is performed through the minimization of the following quantity:
%
\begin{equation}
\label{eq:plane-residuals}
F_{plane}=\sum_{i=1}^{L-1}\sum_{j>i}^{L}\left[(\vec{X_i}-\vec{X_j})\cdot{\vec{k}}-(t_i^*-t_j^*)\right]^2,
\end{equation}
where $L$ is the total number of antennas composing the event, $t_i^*$ and $t_j^*$ are the trigger times for antennas $i$ and $j$ corrected through the cross-correlation treatment, and $\vec{X_i}$, $\vec{X_j}$ their respective positions. The vector $\vec{k}$ can be written as $\vec{k}=\vec{n}$/$v$ where $\vec{n}$ is a vector of norm unity, orthogonal to the wavefront, and $v$ is the wave propagation velocity,  fixed at the value of the velocity of light $c$. The angular coordinates (zenith angle $\theta$ and azimuth angle $\phi$) of $\vec{n}$ are the free parameters of the minimization of $F_{plane}$. \\
\indent As for local background sources, the signal propagates isotropically from the point of emission. In the context of a self-triggering antenna array, we therefore found worth of interest to perform the reconstruction of the direction of the signal assuming a spherical wavefront. This provides a useful way of identifying background sources close to the array. The source location and time of emission of a spherical wave can be estimated by minimizing the following function:
%
\begin{figure}
\includegraphics[width=11cm,height=7cm]{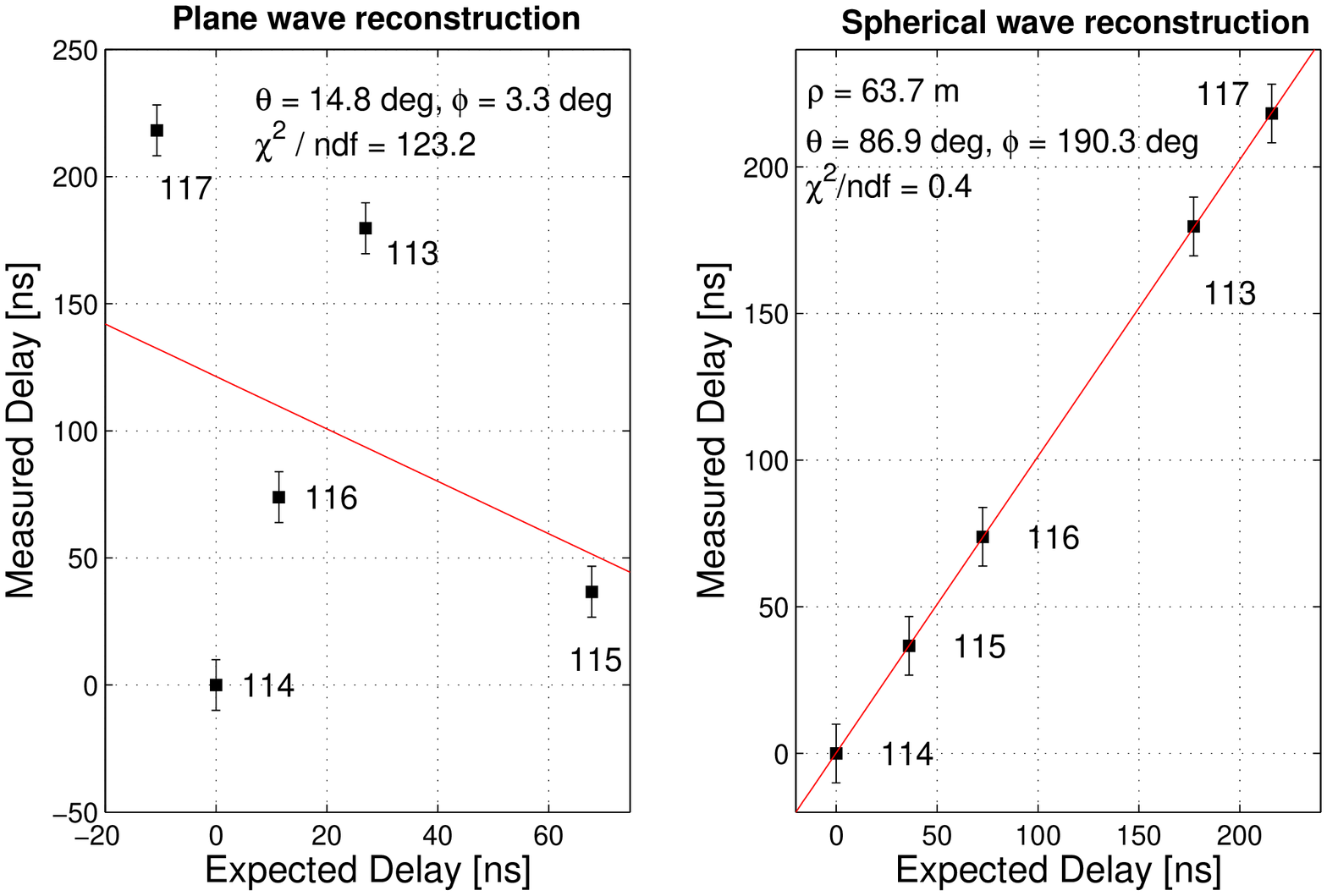} \\
\includegraphics[width=11cm,height=7cm]{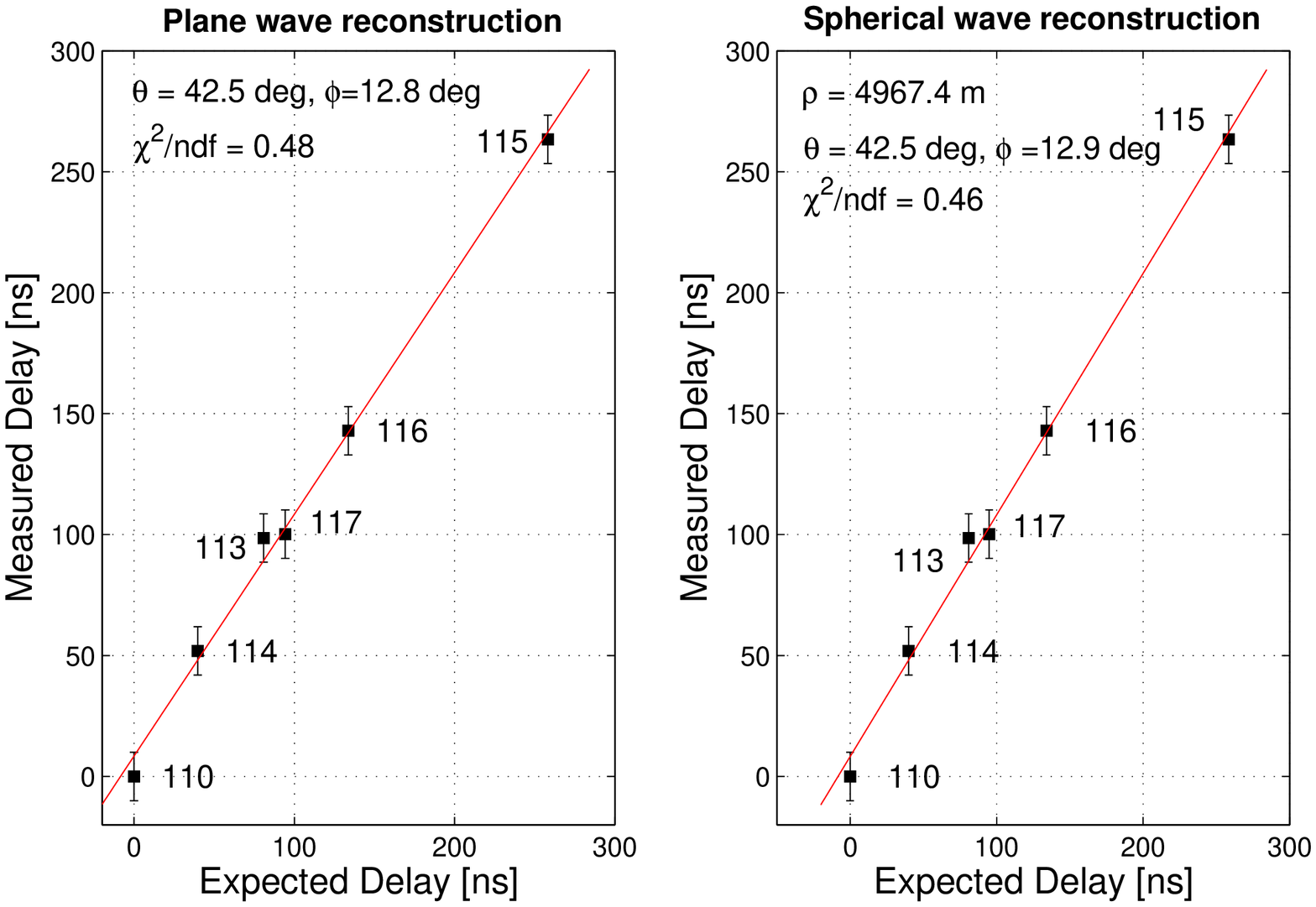}
\caption{ \label{fig:delay-plot} \textit{
  Delay plots for 2 TREND events. Delay plots are defined as the measured trigger times versus the values inferred from plane (left) and spherical
(right) reconstructions. The parameters $\theta$ and $\phi$ are the zenith and azimuth angles of the signal direction of origin. In the case of the spherical reconstruction, the parameter $\rho$ corresponds to the distance to the center of pod 114. Antennas are referred by a number (110, 113 to 117) corresponding to their associated pod number. A resolution of 10~ns is
assumed for the trigger time (see \autoref{sec:reconstruction-performances}). The first triggering antenna is used as the time reference. A linear
fit of the points is performed. The event represented in the two upper plots was recorded during the calibration run performed with a static source placed in the middle of the array (yellow star in \autoref{fig:car-calibration}). The source position reconstructed with the spherical treatment is [2320.3~m, -62.5~m, 2667.6~m] in the cartesian referential defined in \autoref{fig:21cma-layout}, while the true source position is [2322.1~m  -62.0~m 2671.2~m]. Obviously for sources so close to the antennas, the plane reconstruction fails. }}
\end{figure}
%
\begin{equation}
\label{eq:point-residuals}
F_{sph}=\sum_{i=1}^{L}{(t_i^*-t_i^{exp})^2},
\end{equation}
where $L$ is here again the event multiplicity, $t_i^*$ is the corrected trigger time for antenna $i$ and $t_i^{exp}$ its expected value, which can be written as~:
%
\begin{equation}
t_i^{exp}=t_0+\frac{||\vec{X_i}-\vec{X_0}||}{v},
\end{equation}
$\vec{X_i}$ being the antenna position. The source position $\vec{X_0}$ and signal emission time $t_0$ are free parameters of the fit. As for the plane wavefront case, the wave velocity $v$ is taken to be equal to the velocity of light. \\
\indent The plane and spherical reconstructions are compared for each reconstructed event (see \autoref{fig:delay-plot}). The DMNGB routine from the PORT library~\cite{PORT:url} is used for the minimization of the $F_{plane}$ and $F_{sph}$ functions. \\
\indent A visual indication of the quality of the direction reconstruction can be obtained by plotting the measured trigger times versus the values calculated with the reconstruction results (see \autoref{fig:delay-plot}). Points deviating significantly from the first bisector are an indication of a bad reconstruction of the signal direction of origin. A linear fit of this distribution provides a qualitative evaluation of the reconstruction.

\subsection{Reconstruction performances}
\label{sec:reconstruction-performances}
\begin{figure}
\includegraphics[width=6.5cm,height=6.5cm]{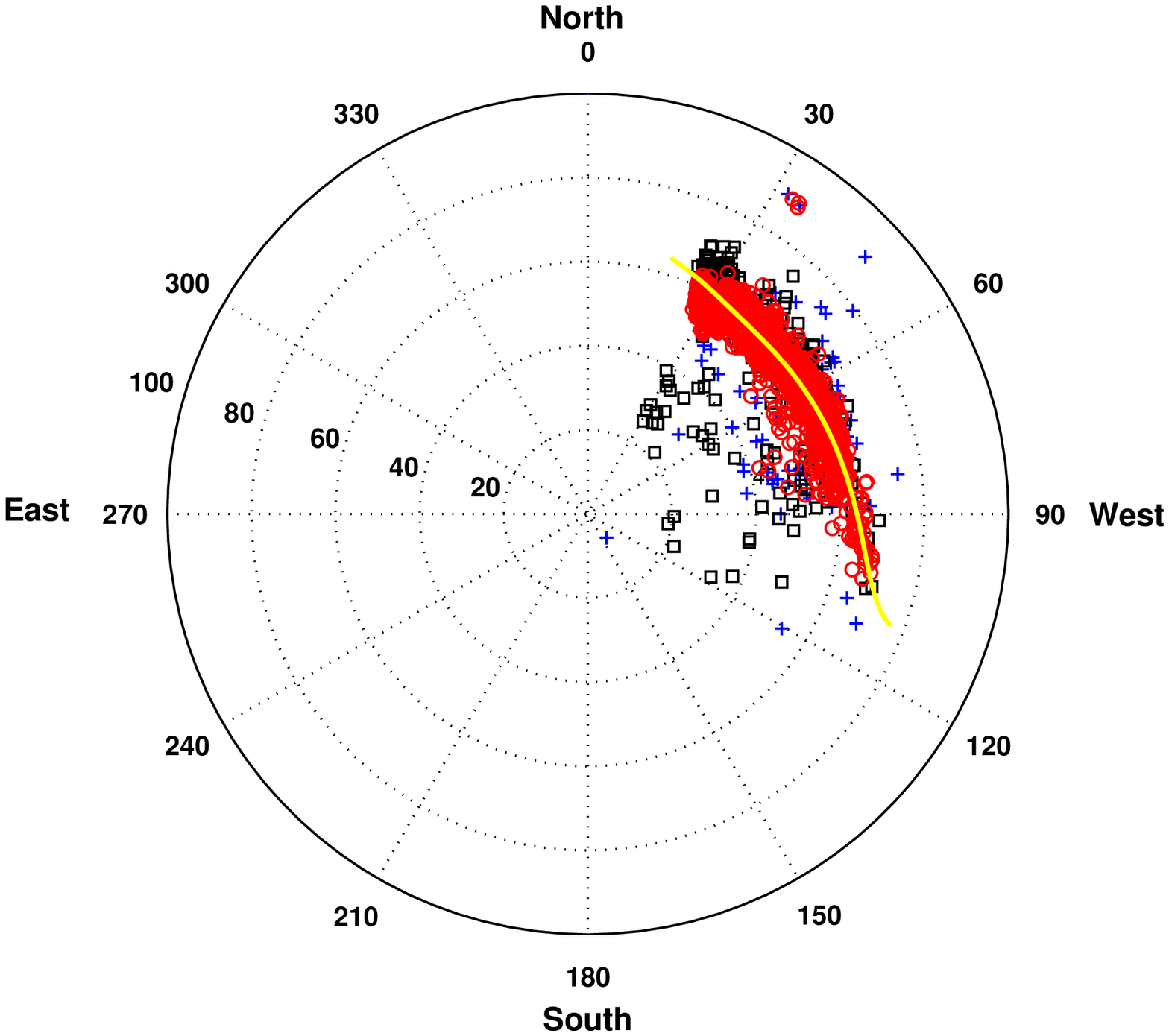}
\includegraphics[width=5.5cm,height=6cm]{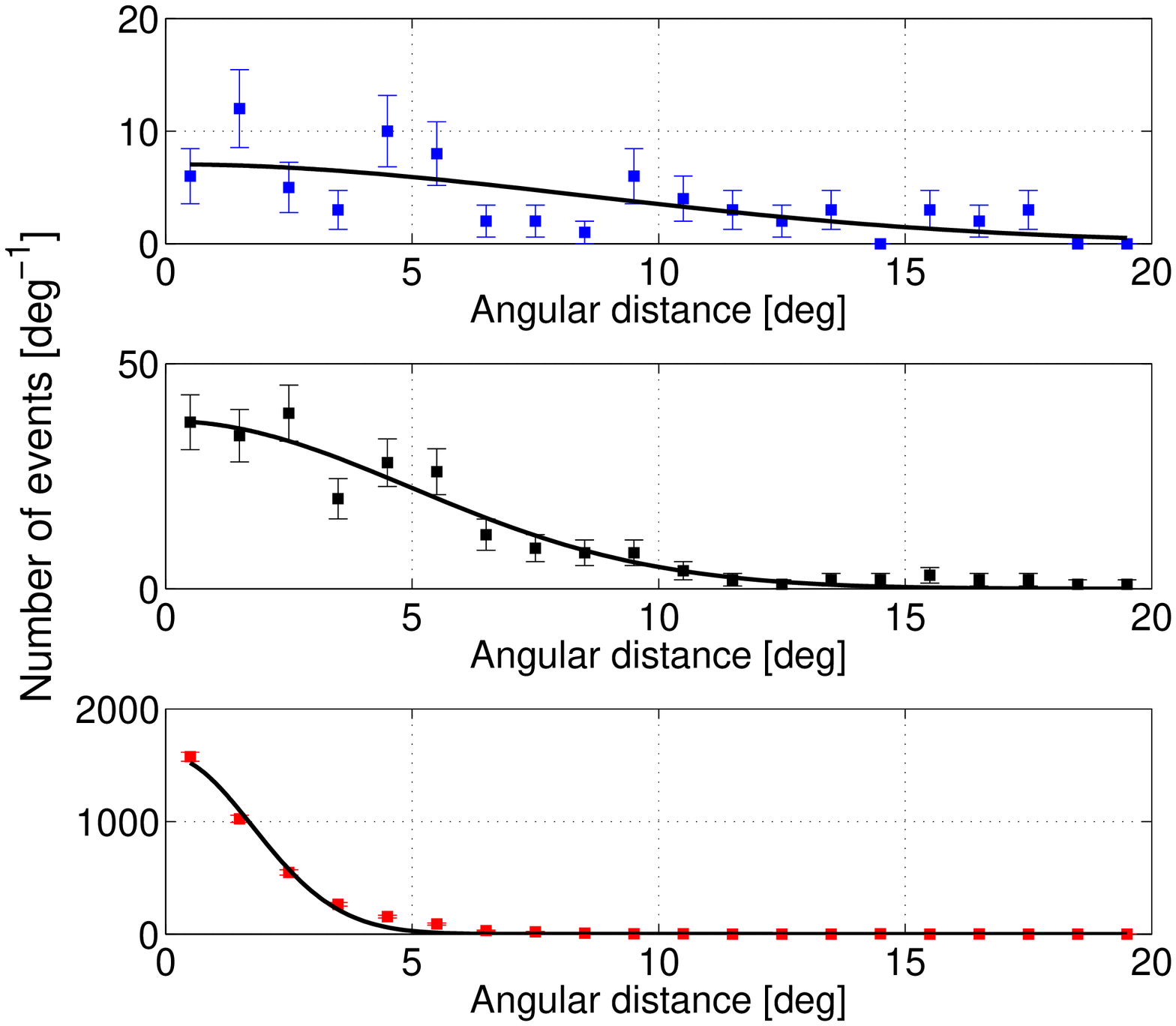}
\caption{ \label{fig:airplane-event} \textit{
  Left: directions of arrival for all events detected within 3 minutes in one of the TREND data sets. The radial and
angular coordinates are respectively the reconstructed zenith and azimuth angles $\theta$ and $\phi$ for the plane wave treatment. Zenith angle values of $\theta$=20, 40, 60, 80 and 100$^{\circ}$ are indicated by labels. The color coding for the events plotted is the same as the one introduced in \autoref{fig:car-calibration}. These events can obviously be associated with a mobile source crossing the sky. Right: histogram of the minimal angular distance to the source trajectory (yellow curve on the left plot) for the same events. The distributions for 4, 5 and 6-antennas events (top to bottom) are well fitted by Gaussian functions with $\sigma$=7.2, 4.4 and 1.5$^{\circ}$ respectively. }}
\end{figure}

\begin{figure}
\begin{center}
\includegraphics[width=5.5cm,height=6.0cm]{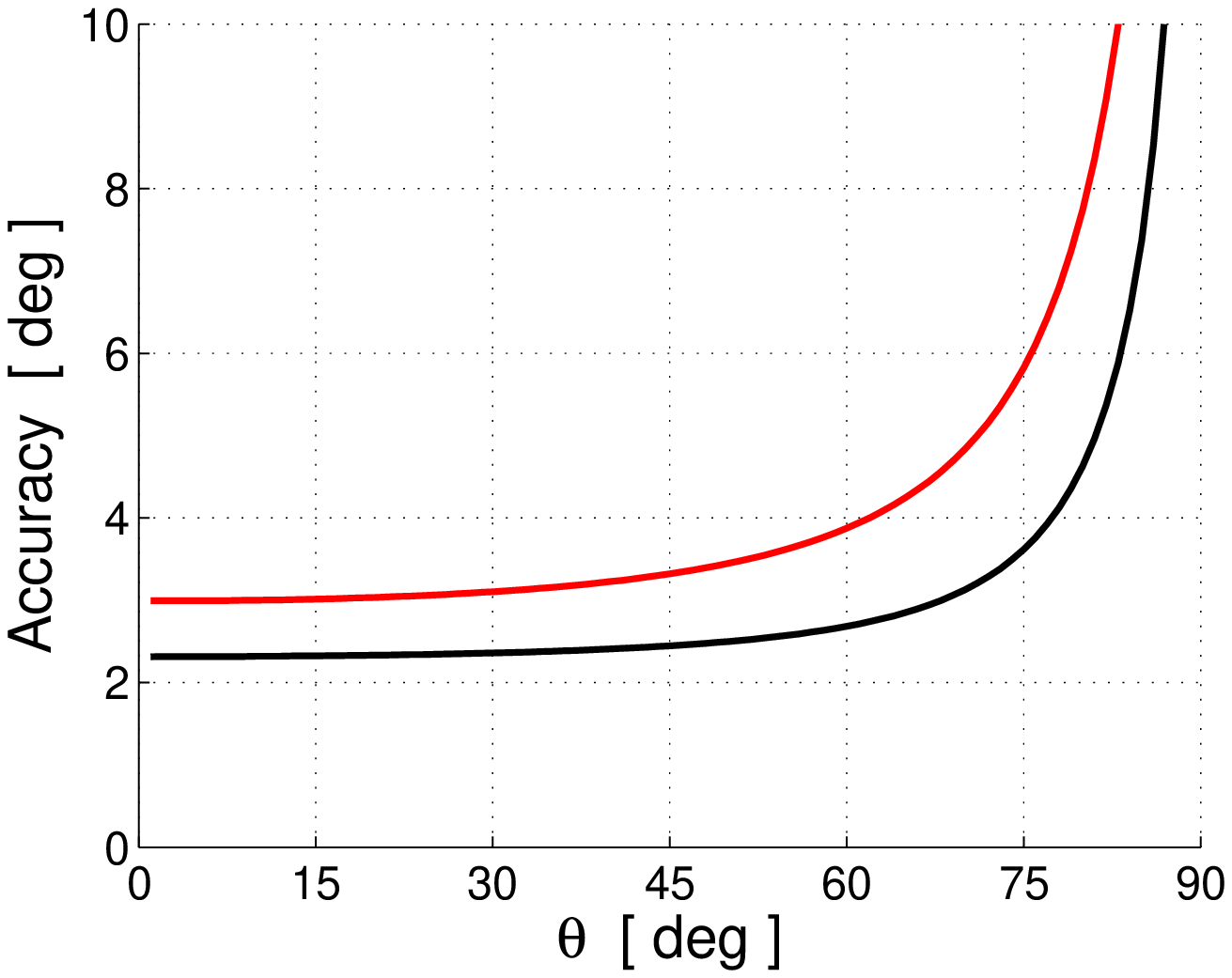}
\includegraphics[width=5.5cm,height=6.0cm]{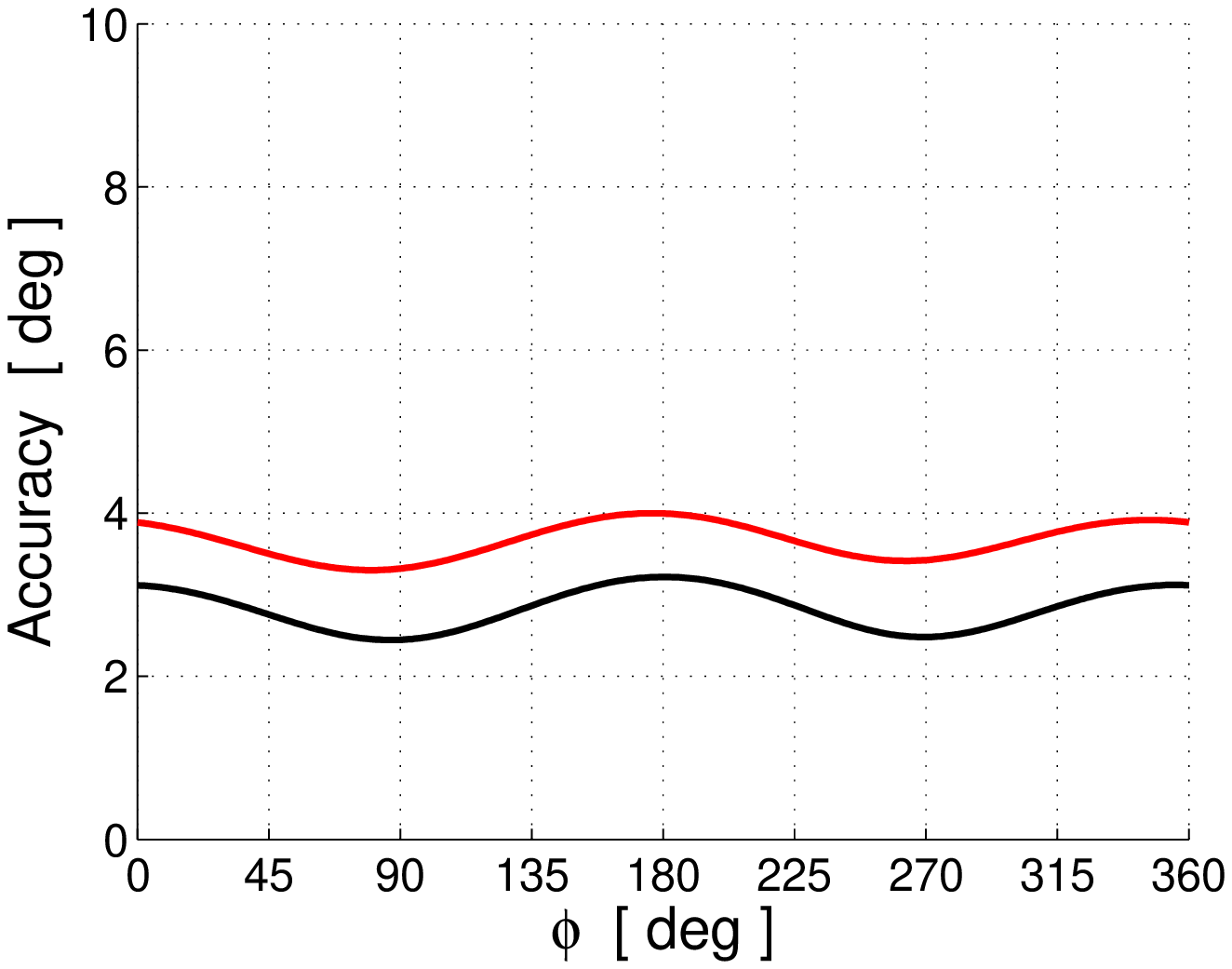}
\caption{ \label{fig:angular-resolution} \textit{
  Theoretical angular resolution for the TREND prototype as a function of the zenith angle $\theta$ (left, $\phi = 90^\circ$) and the azimuth angle $\phi$
(right, $\theta = 45^{\circ}$). Black curves correspond to all 6 antennas triggering. Red curves represent the resolution achieved for events with
antennas 113, 114, 115 and 116 only. A timing resolution $\Delta t = 10$~ns is assumed for this computation.} }
\end{center}
\end{figure}

The angular resolution of the direction reconstruction can be estimated from transient sources crossing the sky. A standard deviation down to 1.5$^{\circ}$ was
achieved on a track observed during 3 minutes in one of the TREND runs (see \autoref{fig:airplane-event}). Several tracks of this type, most likely due to airplanes, were reconstructed in the data recorded with the prototype array, all of them yielding similar values for the angular resolution of the plane wave reconstruction. It should be noted here that the signals generated by this type of sources are likely to be more extended in time than those associated with cosmic ray candidates. It is therefore likely that the trigger time determination is not as precise for these anthropic sources. In this respect, the values of the resolution inferred from these tracks can certainly be considered as conservative. \\
%
\indent The quality of the spherical wave direction reconstruction was also estimated from data taken with a source (in practice, a car) placed in the middle of the array. The source position was measured with a resolution better than 2 meters after correcting the trigger times with the cross-correlation  treatment presented in \autoref{sec:coincidences-selection} (see \autoref{fig:car-calibration}). Sources outside the array were also reconstructed with satisfying precision. The spherical reconstruction of the direction of origin of the events presented in \autoref{fig:airplane-event} for example resulted in an angular standard deviation comparable with the one obtained with the plane reconstruction. \\
\indent In the case of the plane wave treatment, the quality of the direction reconstruction was compared to theoretical expectations assuming Gaussian distributed time errors. The theoretical expectations were computed both analytically and by Monte-Carlo. For the analytical error estimate, we assume leading order error propagation. Details are given in \ref{sec:appendix-A}. For the Monte-Carlo estimate, random values were generated for the wave direction of propagation. The antennas trigger times were then computed for each of these directions, and smeared afterward assuming a Gaussian time resolution. The plane reconstruction algorithm described in the previous section was finally applied to these simulated events, and results were compared to the true source parameters. Assuming a timing resolution of $\Delta t = 10$~ns (corresponding to 2 samples for the digitized data) both treatments yield comparable results for the standard error of the reconstructed direction in the plane wavefront hypothesis, ranging from $2^\circ$ for 6-antennas events up to $8^\circ$ for 4-antennas events, in the worst configuration. These results are in good agreement with the resolution measured with real sources. Note however that a good angular resolution is
achieved for sources associated with small zenith angles only, as the zenith angle determination degrades when moving towards the horizon (see \autoref{fig:angular-resolution}). This effect is due to the coplanar layout of the antennas and could be of some importance in the perspective of neutrino detection, in which case the search focuses on nearly-horizontal EAS.
%

\subsection{Background event rejection \& cosmic ray candidates search}
\label{sec:selection-cuts}
Background rejection is obviously one of the major challenges for a self-triggering experiment like TREND. Even in remote places such as the 21CMA site,
the environment is not totally free from electromagnetic radiations~: electrical transformers, trains or the close-by train station as well as the 21CMA acquisition
room are sources of electromagnetic radiations detected by the TREND setup. These events have to be discriminated from the ones induced by cosmic rays. Fortunately, both types of events exhibit distinct features. \\
\indent Most background sources are located at ground level. In the case of the TREND site, the identified sources are mostly static or follow well-defined trajectories (as it is the case for trains for example). The source intensity is often variable, with bursts periods and quiet states otherwise. The signal duration is likely to be larger than several hundreds of nanoseconds, and a high repetition rate (several pulses within 10 $\mu$s) is frequent. Finally, the corresponding waves are supposed to propagate spherically, with an amplitude decreasing as the inverse of the distance from the point source. \\
\indent Events associated with cosmic rays, on the contrary, follow a random distribution in time and  in direction of origin in a first approximation. The induced antenna signals are generally expected to be very short ($<$200~ns) transient pulses \cite{Ardouin:2005,Horneffer:2006,Huege:2003}, roughly  symmetrical with respect to the position of the signal maximum, and with a subsequent absence of close neighbors in space and time. As discussed already, except for the rare case of vertical showers with energies above $10^{18}$~eV, the radio wavefront can also be approximated by a plane with the TREND prototype setup. Finally, experimental results indicate that the radio amplitude lateral distribution exhibits on the average an exponential decrease~\cite{Allan:1971,Allan:1972}~:
%
\begin{equation}
\label{eq:cosmic-ldf}
A_i=A_0\exp(-\frac{d_i}{d_0}),
\end{equation}
where $A_i$  is the signal amplitude for antenna $i$, $A_0$ is the signal amplitude along the shower axis, $d_0$ the signal attenuation parameter and $d_i$ the distance between the shower axis and antenna $i$.  Values between one hundred and few hundred meters are found experimentally  for $d_0$ \cite{Ardouin:2006,Ardouin:2009a,Falcke:2005,Apel:2009}. \\
\indent Taking into account these various statements, we define a noise rejection procedure for the data analysis. It should be stressed that at the present stage of the TREND project, the priority is to establish the autonomous detection of cosmic rays. The aim of the selection procedure defined in the following is therefore the maximization of the background rejection, even at the cost of a reduced cosmic ray detection efficiency. This efficiency will not be evaluated here.\\
\indent 1) We first exclude all periods of time for which more than three events are reconstructed within three consecutive minutes. These noisy periods correspond to high emission states for background sources. The risk of misidentifying a background event is thus increased during these periods of time, and it is safer to restrict the analysis to quiet periods, corresponding to events rates lower than this threshold. \\
%
\begin{figure}
\includegraphics[width=7cm,height=4cm]{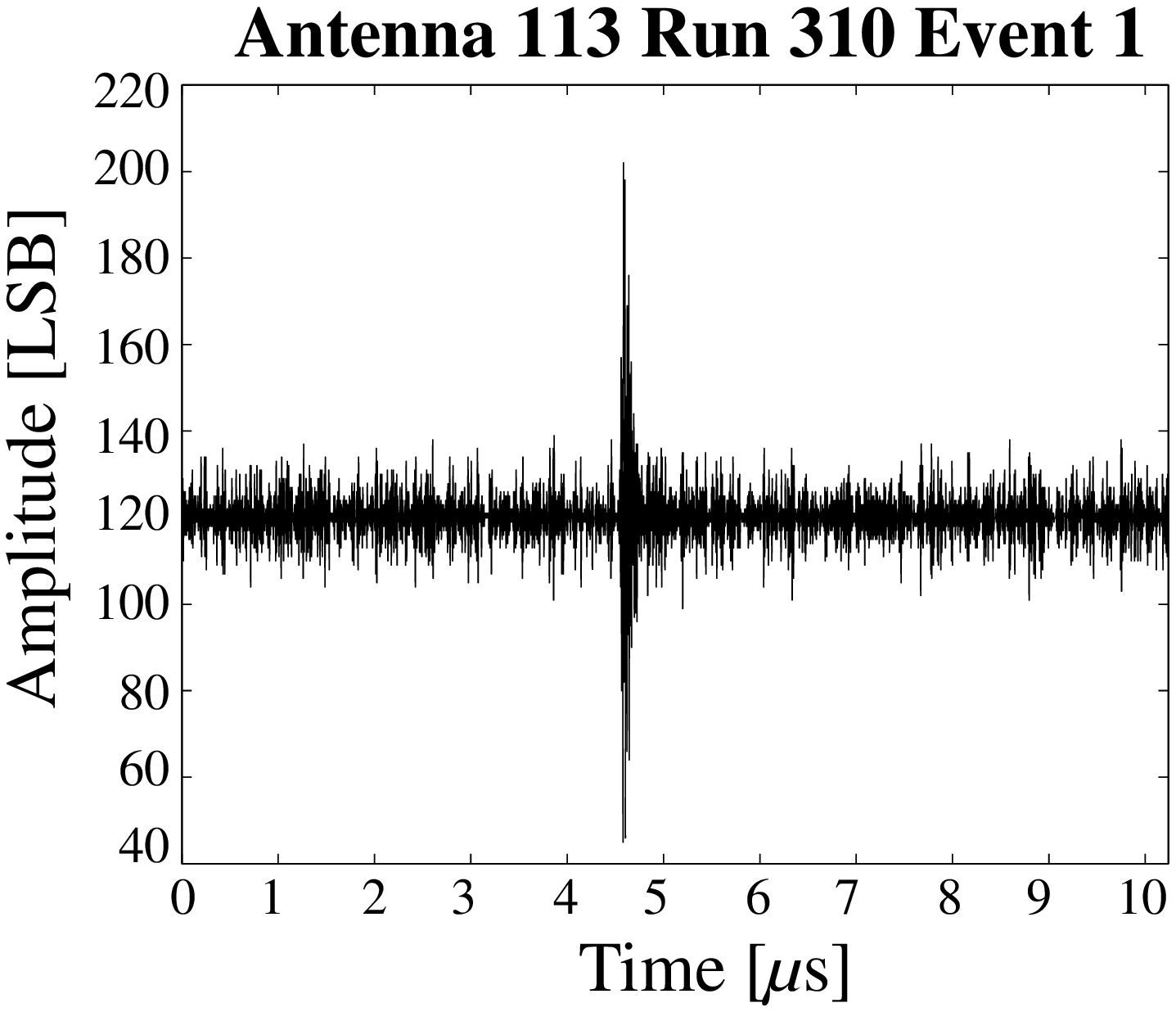}
\includegraphics[width=7cm,height=4cm]{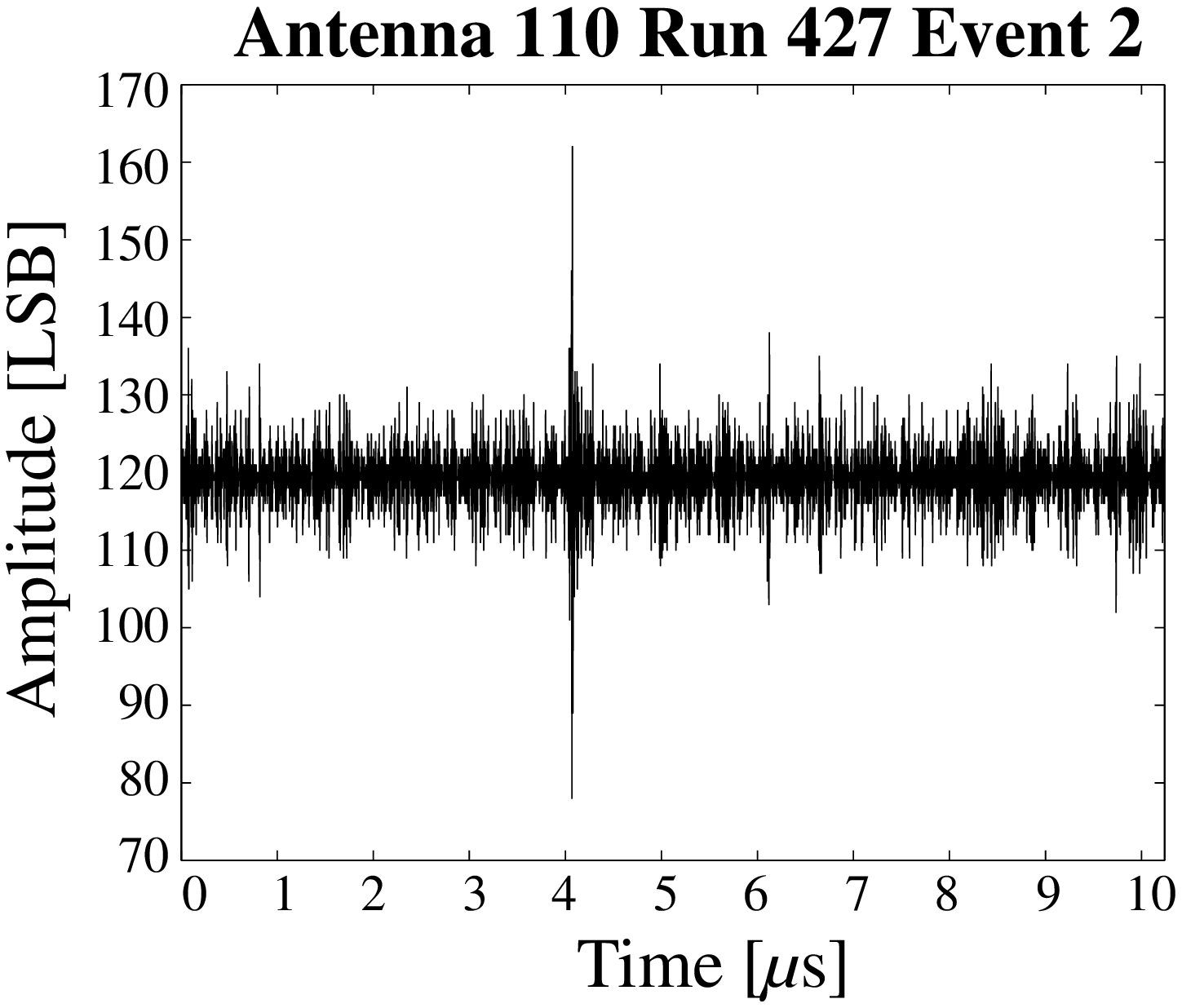} \\
\includegraphics[width=7cm,height=4cm]{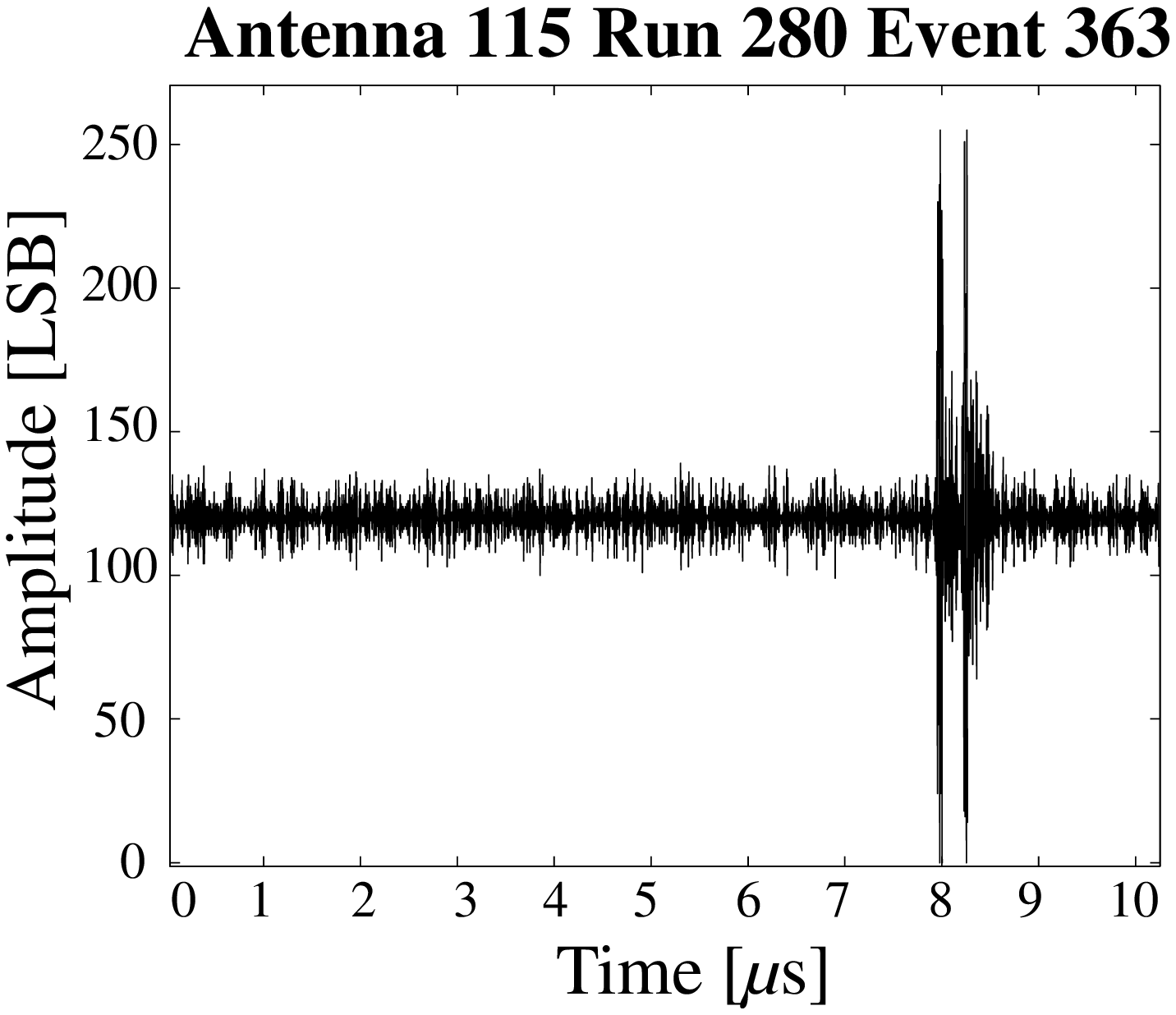}
\includegraphics[width=7cm,height=4cm]{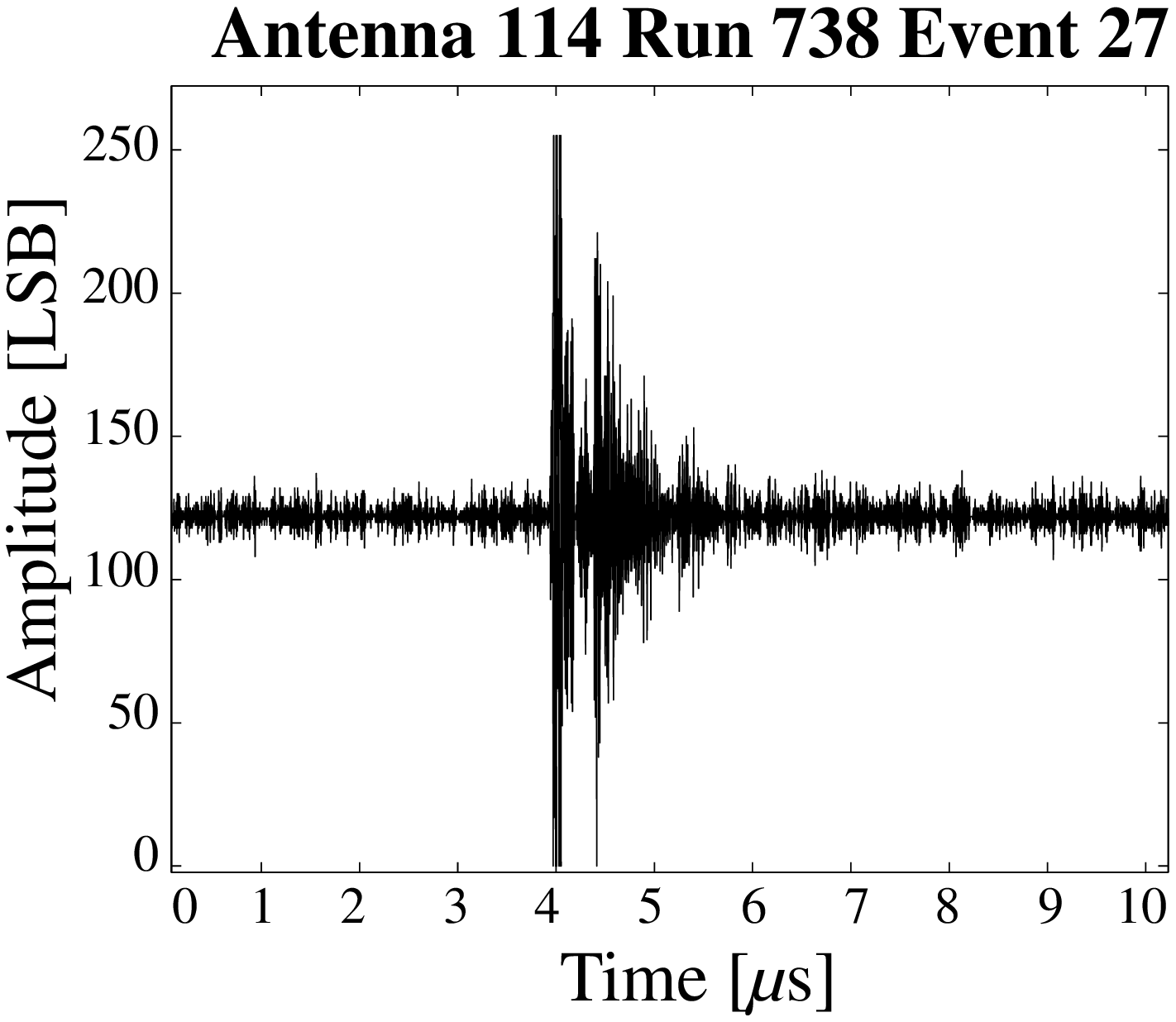}
\caption{ \label{fig:waveform-samples} \textit{ Top: two waveforms passing rejection criterion 2. Bottom: 2 waveforms rejected with cut 2. Time is expressed here in microseconds. }}
\end{figure}
\indent 2) As transients associated with EAS are expected to last for 200~ns at most, all signals with amplitudes significantly higher than the average antenna noise level for more than 400~ns  are discarded, as well as waveforms with several transients in the 10~$\mu$s sampled window. This cut is illustrated through some examples in \autoref{fig:waveform-samples}. \\
%
\begin{figure}
\begin{center}
\includegraphics[width=8cm,height=8cm]{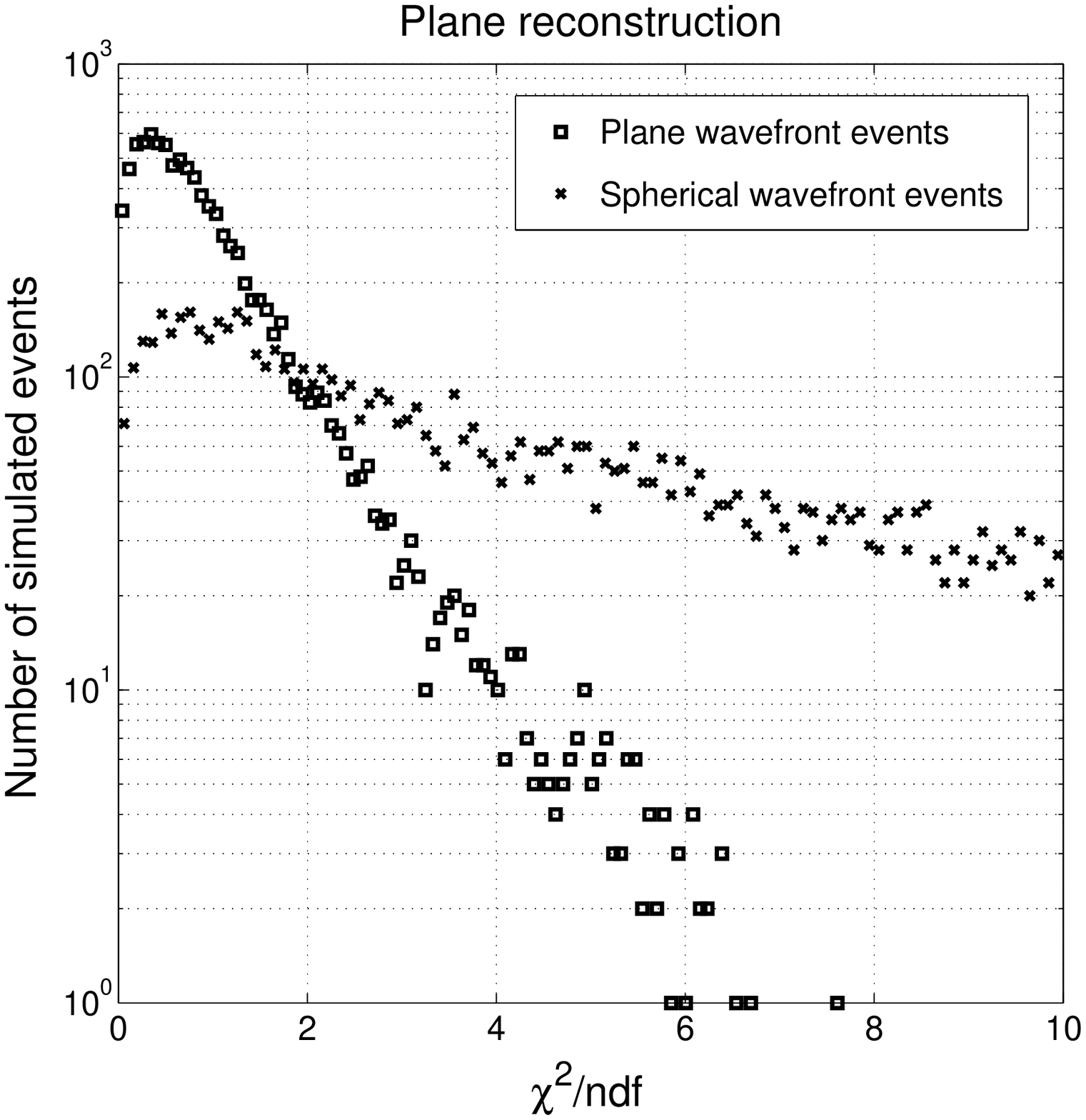}
\includegraphics[width=8cm,height=8cm]{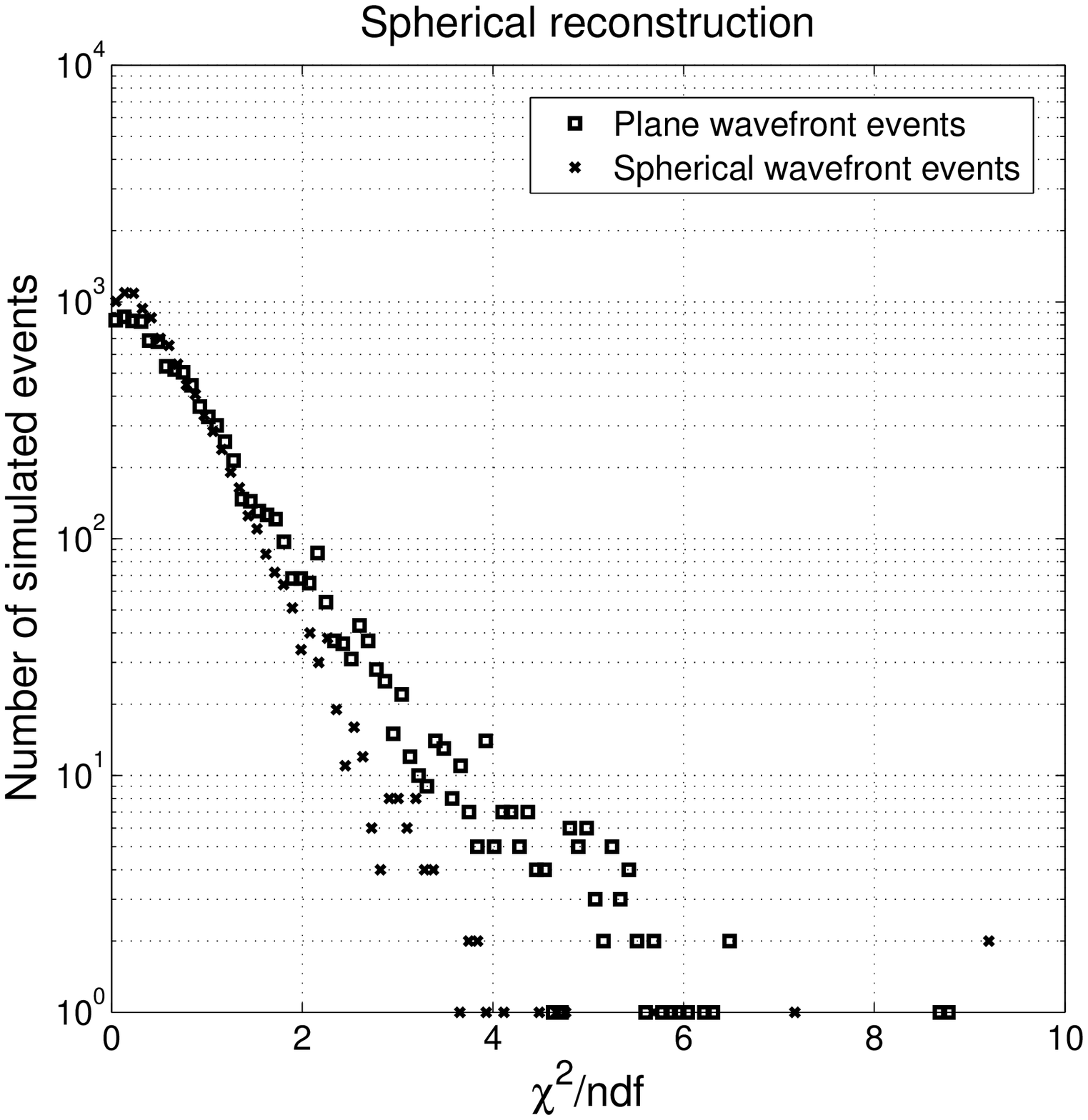}
\end{center}
\caption{ \label{fig:delay-plot-chi2} \textit{ Top~: histogram of the $\chi^2/ndf$ values obtained from the linear fit of the delay plots
(see \autoref{fig:delay-plot}) for the plane wave reconstruction of 10000 simulated plane wavefront events (squares) and 10000
simulated spherical wavefront events (crosses). In the case of spherical waves, the wave source
position is randomly chosen on the ground within a distance of 500~m from the center of the antenna array. Bottom~: $\chi^2/ndf$ distribution for the
spherical wave reconstruction of the same events. }}
\end{figure}
\indent 3) The accepted events are required to be well reconstructed. The quality of the direction of origin reconstruction for a given event can be quantified by the $\chi^2$/ndf value of the linear fit of the delay plot (expected trigger times versus measured values, see \autoref{fig:delay-plot}). This was studied with Monte-Carlo simulations of events associated with plane and spherical wavefronts. The simulation of events associated with plane wavefronts is described in \autoref{sec:reconstruction-performances}. The simulation of events associated with spherical wavefronts proceeds as follows~: a source position is randomly chosen within 500~m from the center of the TREND prototype. The antennas trigger times are then computed assuming a spherical propagation of the signal from the point source and -as for the plane wavefront case- the computed trigger times are then randomly smeared, assuming a Gaussian distribution with $\sigma$ = 10~ns for the timing resolution. \\
\indent 10000 plane wavefront events and 10000 spherical wavefront events were simulated as triggering the six antennas of the prototype. The plane and spherical wavefronts reconstruction procedures described in \autoref{sec:reconstruction} were then applied to these simulated events. A linear fit was performed on their delay plots (see \autoref{fig:delay-plot}). As expected, the $\chi^2$/ndf distributions are similar for both types of events for the spherical reconstruction, while the plane reconstruction for spherical wavefront events is, on the average, worse than for plane wavefront events (see \autoref{fig:delay-plot-chi2}). It seems reasonable from this result to exclude all events associated with a $\chi^2$/ndf value above 2 for the plane reconstruction. However, data tend to exhibit larger $\chi^2$/ndf values than what is observed with simulations. This is most certainly due to the systematic uncertainties affecting the measurement of the signal delay in the coaxial cables and optical fibers, estimated at $\pm$0.5 sample (i.e. 2.5~ns) for each antenna. The cut set to the plane waveform reconstruction of the direction of origin in our cosmic ray search procedure is therefore~:
$$\chi^2/ndf<5.$$

\indent 4) Since most background sources are expected to be located at ground level, the signal/noise ratio should increase when moving towards the zenith. In order to establish the autonomous radio-detection of cosmic rays, we therefore choose to limit our candidates search to events reconstructed with zenith angle values below $65^{\circ}$~:
$$\theta_{plane}<65^{\circ}.$$
%
%
\indent 5) It has been shown in \autoref{sec:reconstruction} that the time resolution and array extension of the TREND prototype do not allow for the measurement of the curvature radius of the EAS radio wavefront, except for the very rare case of nearly vertical showers with energies above $10^{18}$~eV. Both plane and spherical wavefronts treatments should therefore yield similar results for the reconstructed direction of origin of the signal. We therefore request the following conditions for the results of the direction of origin reconstruction~:
$$\sqrt{(\theta_{sph}-\theta_{plan})^2+(\phi_{sph}-\phi_{plan})^2}<d_{cut},$$
where $d_{cut}$ = 10, 7 and 4$^{\circ}$ for 4, 5 and 6-antennas events respectively. These values roughly correspond to the quadratic sum of the average error for the spherical and plane wave reconstructions. \\
\indent These various selections were applied to simulated data in order to estimate the background rejection efficiency. This Monte Carlo simulation was produced following the method described in \autoref{sec:reconstruction-performances}, with ground source positions randomly chosen within 10000~m from the array center. Over $10^5$ simulated events, none passes cuts 3 to 5. This is an interesting result, even though it has a limited validity only,  as the simulation performed here presents very basic characteristics, far from reality. \\
\indent 6) The amplitude information should in principle provide a powerful lever for background rejection~: as pointed out already, the lateral profile of the signal amplitude should preferentially decrease exponentially for an event associated with an EAS (see \autoref{eq:cosmic-ldf}). The combination of this rapidly decreasing signal amplitude with a large curvature radius of the wavefront could constitute a distinct signature of cosmic ray candidates as demonstrated recently \cite{Ardouin:2006}. Indeed, background events should either be associated with large curvature radii and roughly constant amplitudes in the case of distant sources, or spherical wavefronts and rapidly decreasing amplitudes for close sources. It is therefore believed that background rejection efficiency of an antenna array will increase with its size. The  reduced extension of the 6-antennas TREND prototype array ($\sim$200~m in its largest dimension) and the limited precision of the calibration technique presently used (see \autoref{sec:antenna-sensitivity}) do not allow for a reliable reconstruction of the shower core position and lateral amplitude profile. This complete reconstruction of the shower characteristics for events passing cuts 1 to 5 will therefore be performed (see \autoref{sec:results}), but not included in the cosmic ray candidate search procedure. The minimization function for the shower geometry reconstruction is assuming an exponential decrease of the lateral amplitude and is given by~:
%
\begin{equation}
F_{shower}  = \sum_{i=1}^{L}(s_0-a_0\times{d_i}-s_i)^2,
\end{equation}
where the sum runs over all triggering antennas, $a_0$ is given by $a_0=\frac{20}{\log(10)\times{d_0}}$, $d_0$ being the attenuation parameter of the exponential fit of the lateral amplitude profile as defined in \autoref{eq:cosmic-ldf}. $s_0$ is the electromagnetic field amplitude in decibels along the shower axis, $d_i$ the distance from antenna $i$ to the shower axis and $s_i$ the measured antenna amplitude in decibels. The parameters $s_0$, $a_0$, and the shower core position ($x_0$, $y_0$) are free parameters of the minimization. The shower direction is given by the vector $\vec{n}$ previously determined through the plane wave direction reconstruction of the event (see \autoref{eq:plane-residuals}). As before, the function minimization is performed with the DMNGB routine from the PORT library \cite{PORT:url}.

\subsection{Results}
\label{sec:results}
\begin{figure}
\includegraphics[width=6.5cm,height=6.5cm]{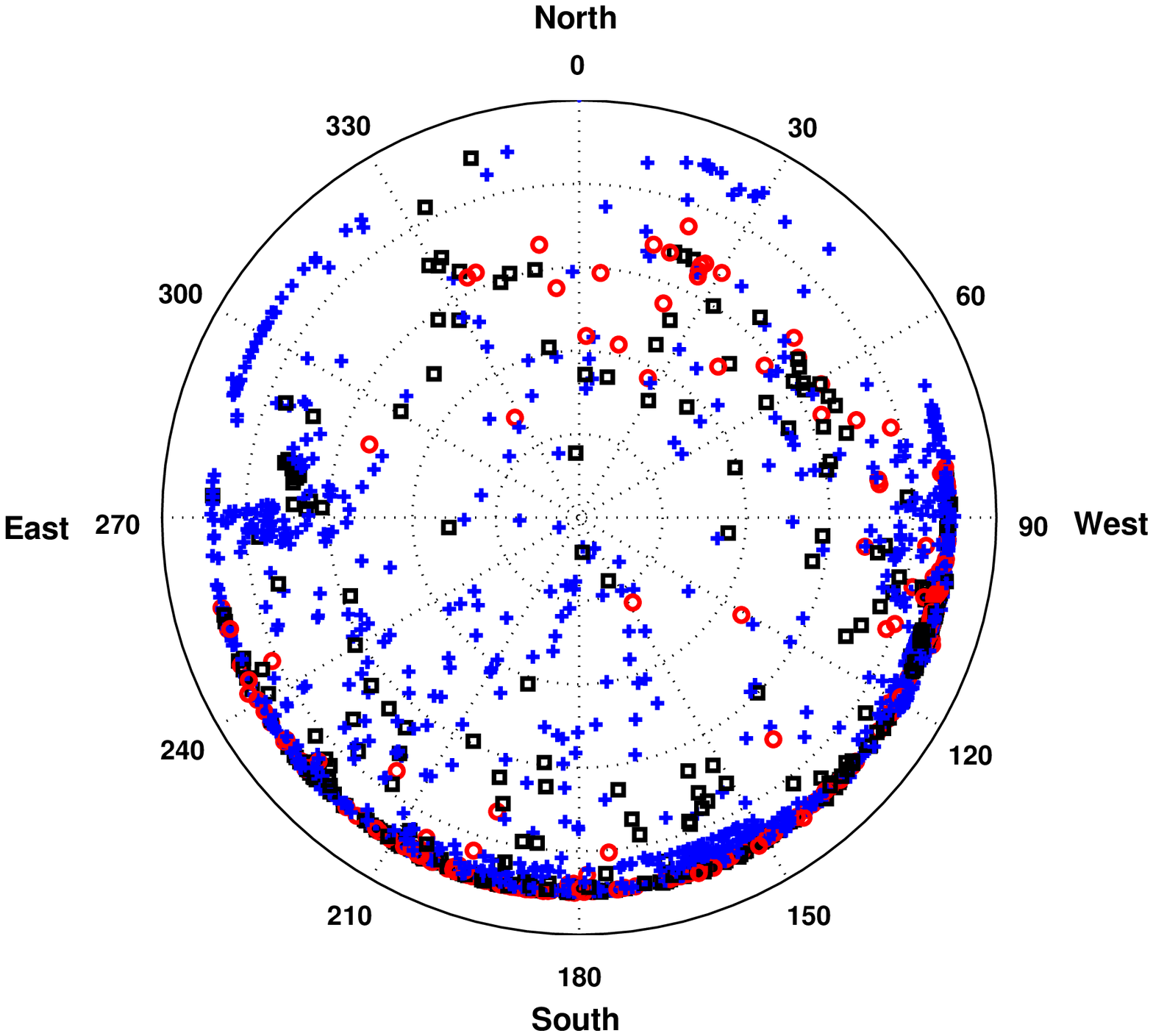}
\includegraphics[width=6.5cm,height=6cm]{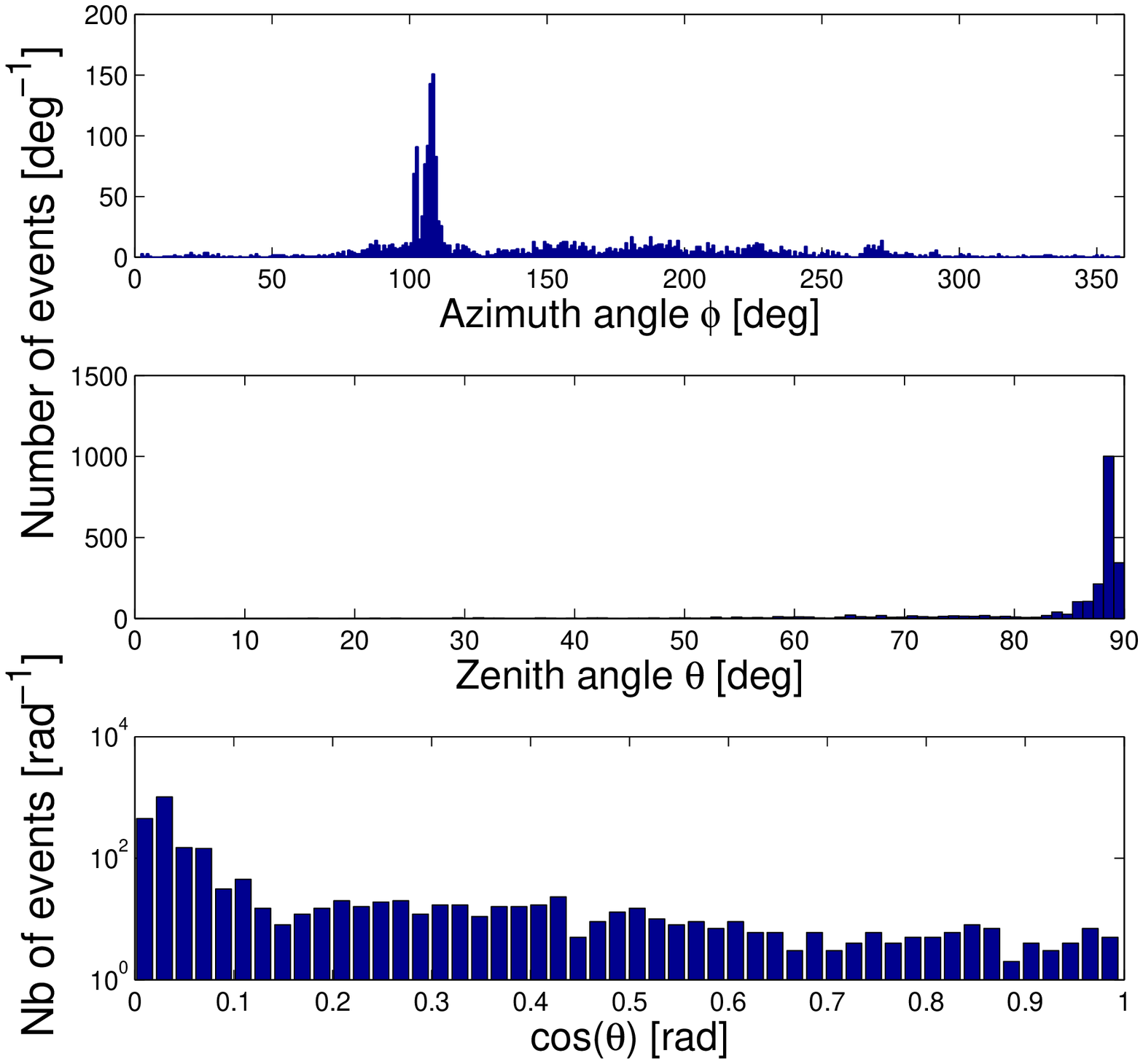}
\caption{ \label{fig:quiet-skyplot} \textit{
  Left : plot of the direction of origins for the 2275 events reconstructed during quiet periods. The color coding for the events plotted is the same as the one introduced in \autoref{fig:car-calibration}. Right: histogram of the reconstructed azimuth (top) and zenith (middle) angles for the same events by bins of 1$^{\circ}$. Also shown in the bottom plot is the $\cos \theta$ distribution for these events by bins of 0.02 radians. The direction $\phi\sim$105~$^{\circ}$ corresponds to 2 electrical transformers close to the 21CMA acquisition room. }}
\end{figure}

In this section, we apply the selection procedure defined in the previous section to the data recorded in 2009 with the 6-antennas TREND prototype. \\
\indent The setup ran for 584.7 live hours in 2009. The noisy periods amount to 31\% of this total. In the 403.4 hours selected as quiet periods, 2275 events are successfully reconstructed  (see \autoref{fig:quiet-skyplot}). The vast majority of these events lie along the horizon and are mostly associated with background sources. It is noticeable that few of these ground events are reconstructed in the North direction. This can be explained by the fact that no potential background sources can be found North from the array, while several exist in the South (railroad), East (train station) or West (DAQ room and electrical transformers) directions. \\
%
%
\begin{figure}
\begin{center}
\includegraphics[width=6.5cm,height=6cm]{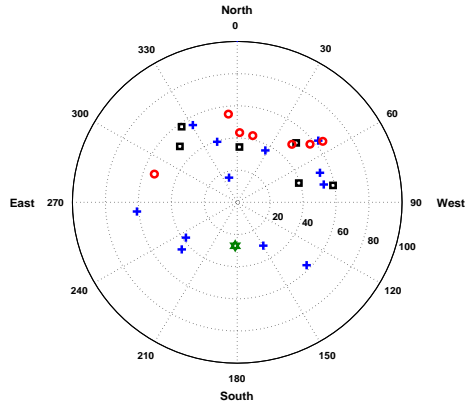}
\end{center}
\caption{ \label{fig:candidates-skyplot} \textit{
  Repartition of the direction of arrival for the 25 cosmic ray candidates detected with the TREND 6-antennas prototype. Zenith angle values of $\theta$=20, 40, 60, 80 and 100$^{\circ}$ are labeled. The color coding for the events plotted is the same as the one introduced in \autoref{fig:car-calibration}. Also shown as a green star is the direction of the geomagnetic field at the TREND location. }}
\end{figure}
\indent After applying cuts 2 to 5 to the 2275 events recorded in quiet periods, only 37 events are left. A detailed analysis ends in a further rejection of 12 of these candidates, which are possibly related to other events reconstructed close in time or space. On the contrary, all 25 surviving candidates are clearly isolated in time and space, and are therefore considered as cosmic ray candidates. We plot in \autoref{fig:candidates-skyplot} the reconstructed arrival directions for these events. It is worth noting that 20 of them lie in the Northern part of the sky, while there are significantly less background sources at ground level towards North. As various biases have not been taken into account in the present analysis (in particular the antenna lobe), it is certainly premature to relate the observed North-South asymmetry to a possible geomagnetic origin of the radio signal, as done by CODALEMA \cite{Ardouin:2009a}. \\
%
\begin{figure}
\includegraphics[width=7cm,height=6cm]{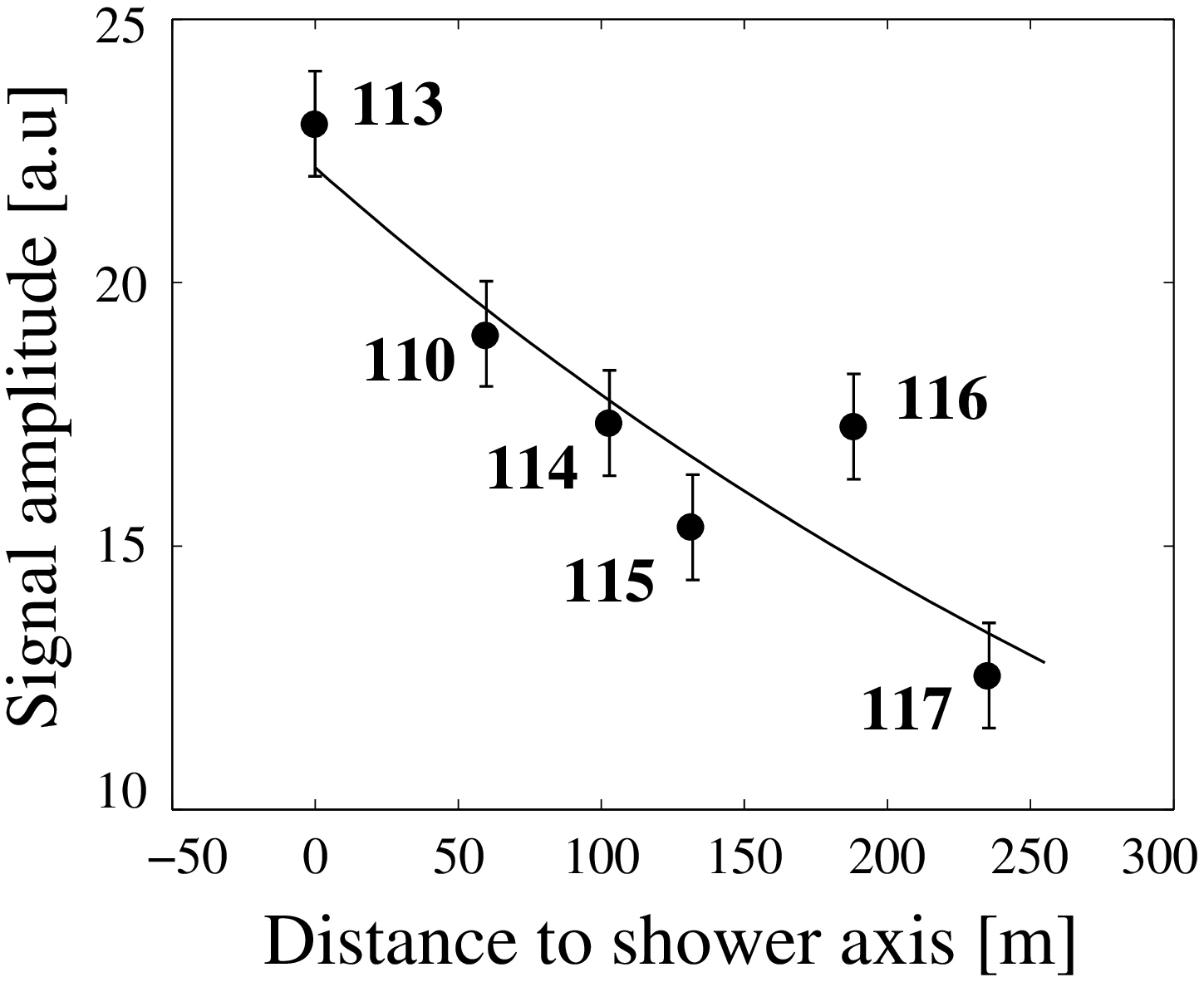}
\includegraphics[width=5cm,height=6cm]{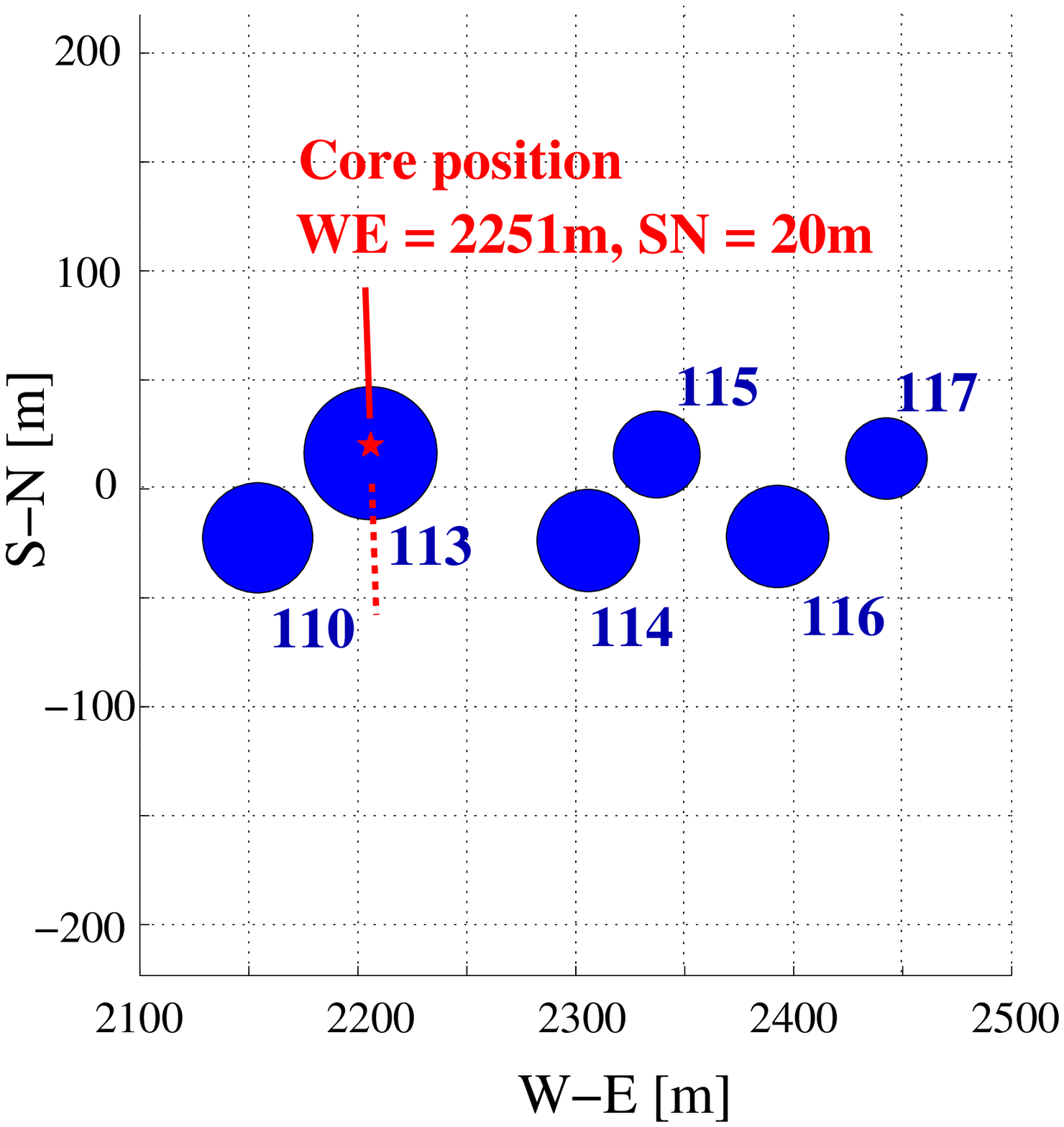}  \\
\includegraphics[width=7cm,height=6cm]{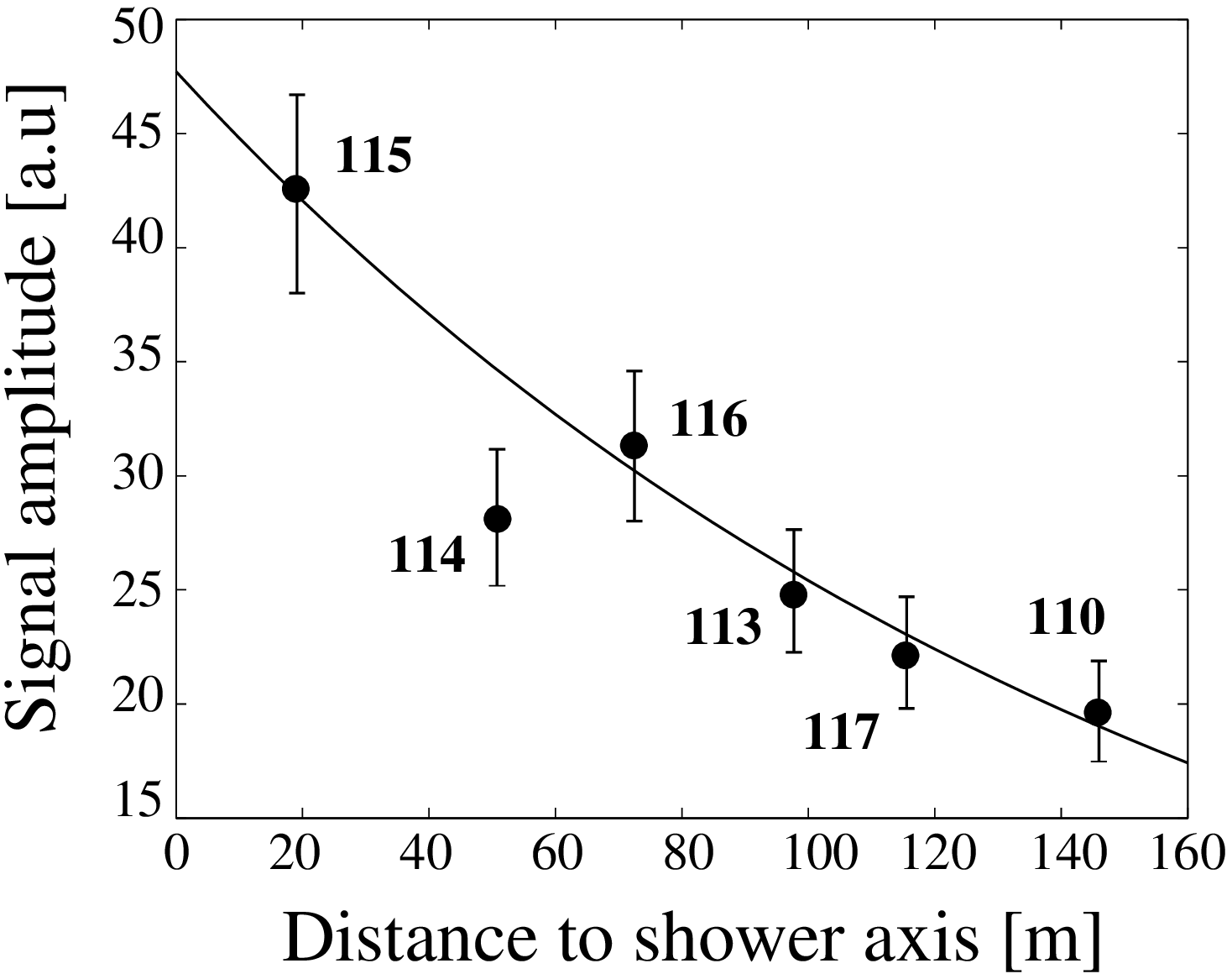}
\includegraphics[width=4.5cm,height=6cm]{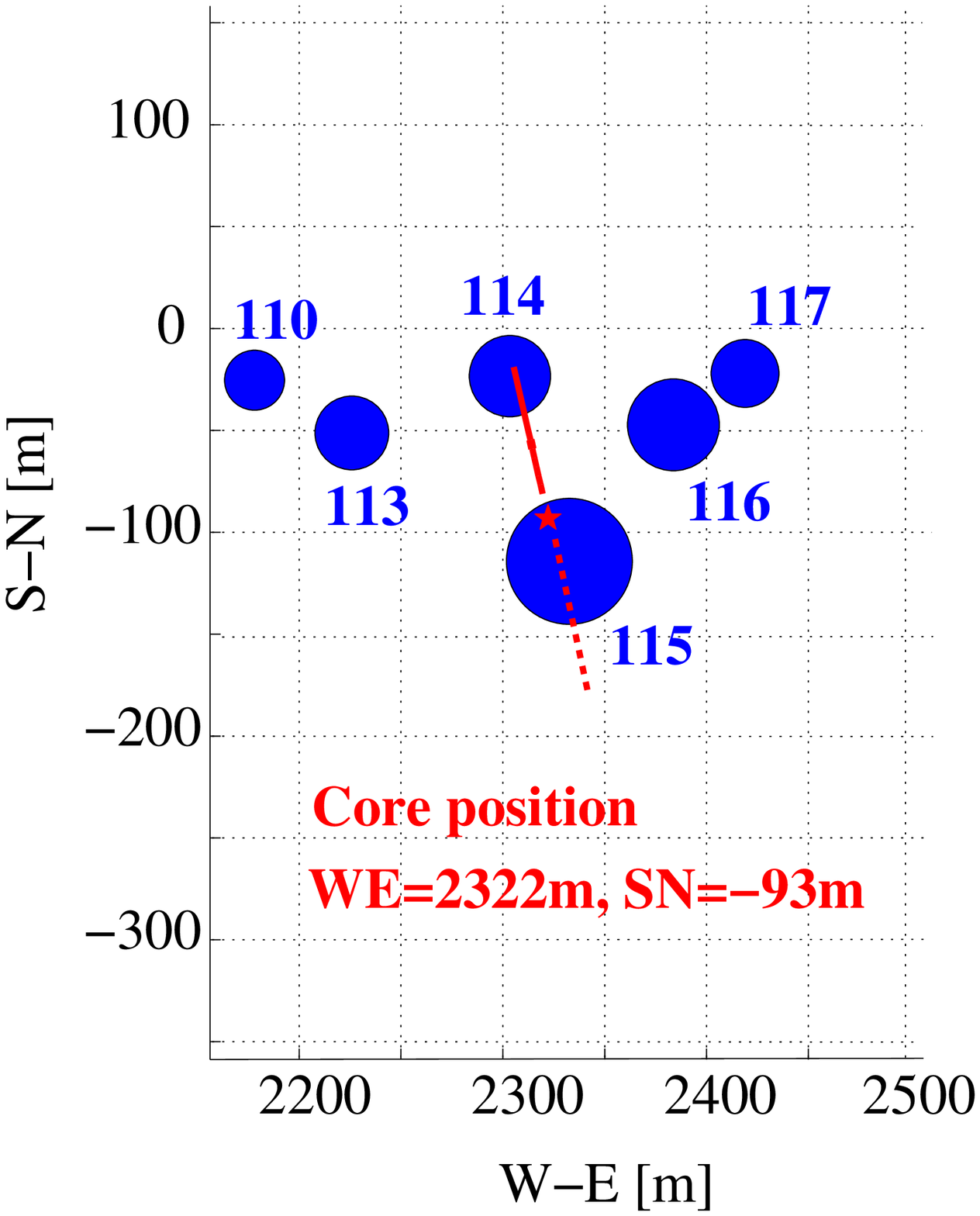}  \\
\caption{ \label{fig:candidates-ldf} \textit{
  Left~: signal amplitude as a function of antenna distance to the shower axis and exponential fit for 2 cosmic ray candidates. Only statistical
uncertainties on the amplitudes are shown.  For the second candidate, the amplitude of antenna 114 is not taken into account~: as the antenna was
placed atop a 5~m high pole at this time, its relative calibration is not reliable. Right~:  projection of the reconstructed shower axis (red line, dashed is underground) on the detector plane. The core position is represented by a  red star. A disk of radius proportional to the signal amplitude is drawn around each antenna
position. Note that these 2 events were recorded with 2 different antennas layouts. The shower plane directions are ($\theta=44 \pm 8^{\circ}$, $\phi=2 \pm 2^{\circ}$) and ($\theta=42 \pm 5^{\circ}$, $\phi=13 \pm 3^{\circ}$) respectively. Values  $d_0$=462~m and $d_0$=158~m are found for the  exponential fit of the lateral amplitude profile (see \autoref{eq:cosmic-ldf}).}}
\end{figure}
\indent A complete reconstruction of the shower characteristics was successfully performed on 18 of these 25 candidates, as shown in \autoref{fig:candidates-ldf} for two nice examples. The average value for the attenuation parameter is $\left<d_0\right> \simeq 200$~m, a result consistent with calculated values~\cite{Huege:2005} and available radio-detection data~\cite{Ardouin:2006, Apel:2009}. It should be stressed however that in these experiments, ground detectors are used to trigger the antenna array.

\subsection{Validation of the cosmic ray search procedure}
\begin{figure}
\begin{center}
\includegraphics[width=4cm,height=7cm]{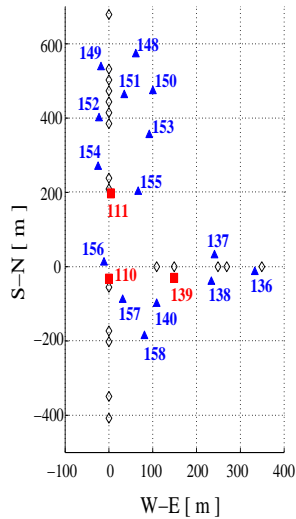}
\end{center}
\caption{ \label{fig:crosspoint-layout} \textit{
  Layout of the TREND setup installed at the cross-point of the two 21CMA baselines. Antennas are shown as blue triangles, scintillators as red squares. Also shown as black diamonds are the 21CMA pods. TREND detectors are labeled by numbers corresponding to the pods they are connected to. }}
\end{figure}

\begin{figure}
\begin{center}
\includegraphics[width=6.5cm,height=6.5cm]{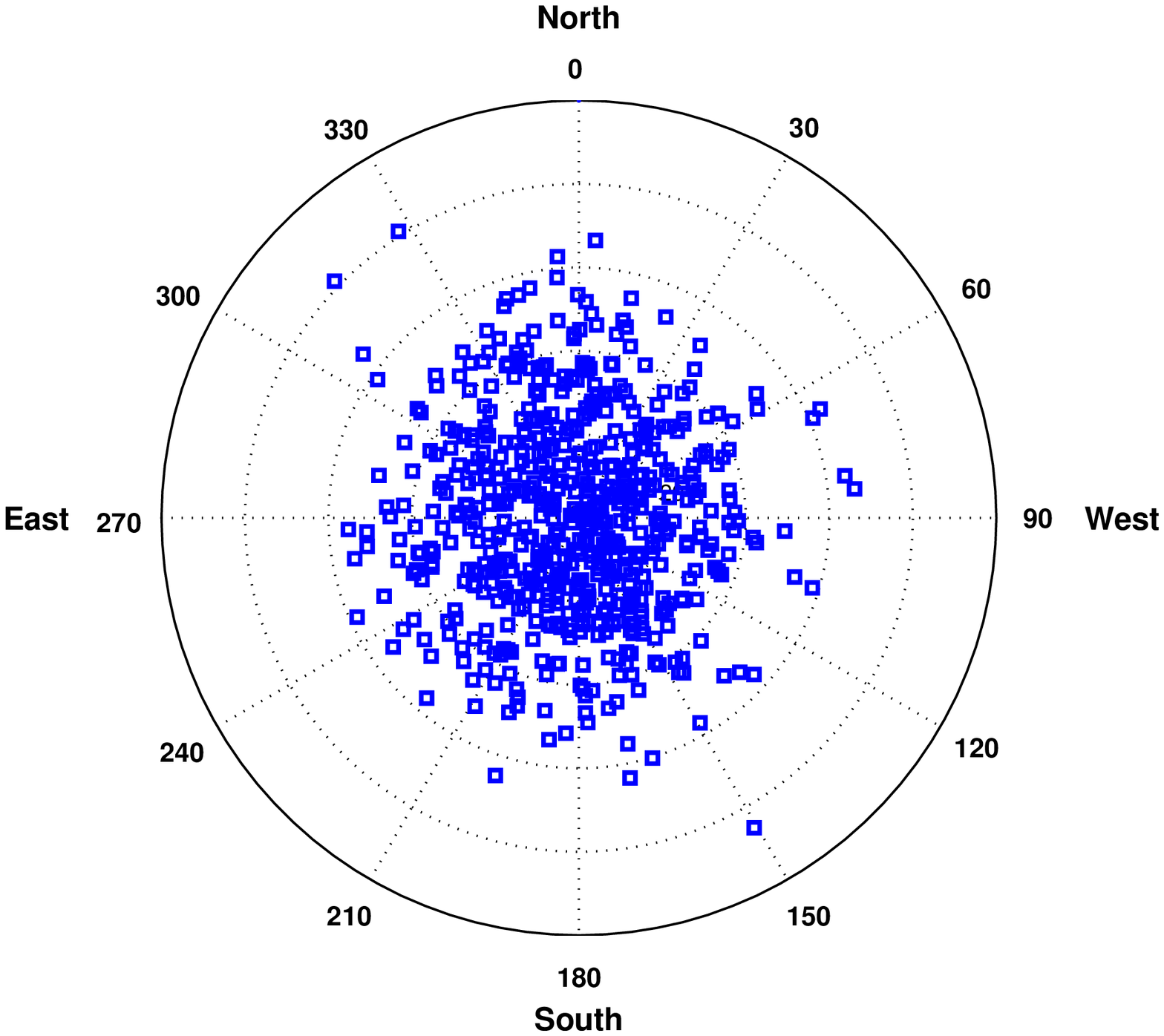}
\includegraphics[width=6.5cm,height=6.5cm]{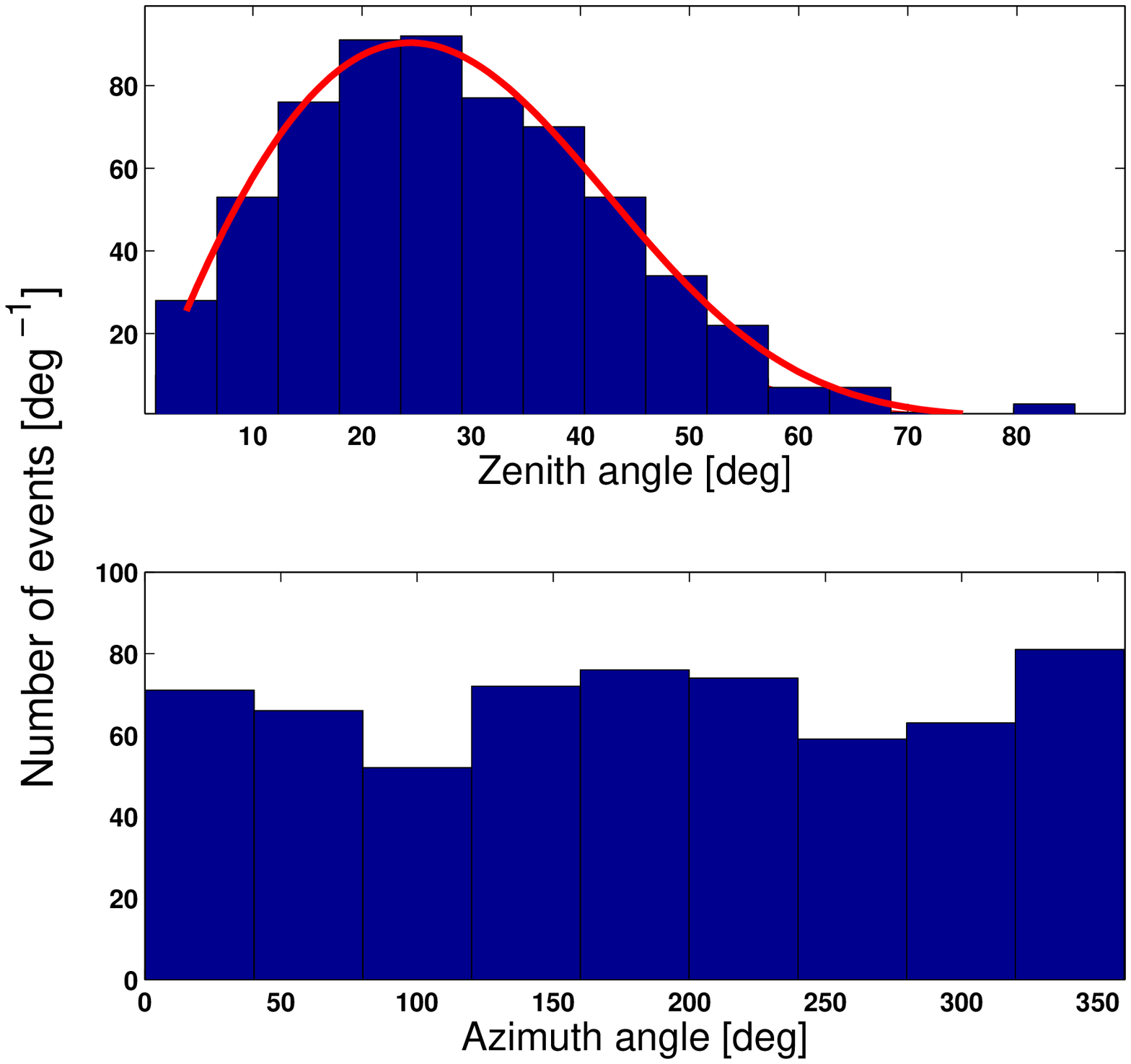}
\end{center}
\caption{ \label{fig:scints} \textit{
  Left: repartition of the direction of arrival for the 620 3-fold events recorded with the scintillator array. Right: distribution of zenith and azimuth angles for the same data by bins of 6 and 40$^{\circ}$ respectively. The zenith angle distribution is fitted by a curve given by $\frac{dN}{d\theta} = (a+b\theta ) \sin \theta \cos \theta \times \frac{1} {1+e^{(\theta - \theta_0 )/ \theta_1 }}$, corresponding to the expected zenith distribution for a scintillator array~\cite{Ardouin:2009a}. The adjusted values for a, b, $\theta_0$ and $\theta_1$ are respectively 68$^{\circ -1}$, -1, 45$^{\circ}$ and 12$^{\circ}$. These are standard values given the angular resolution of the array.}}
\end{figure}

\begin{figure}
\begin{center}
\includegraphics[width=16cm,height=12cm]{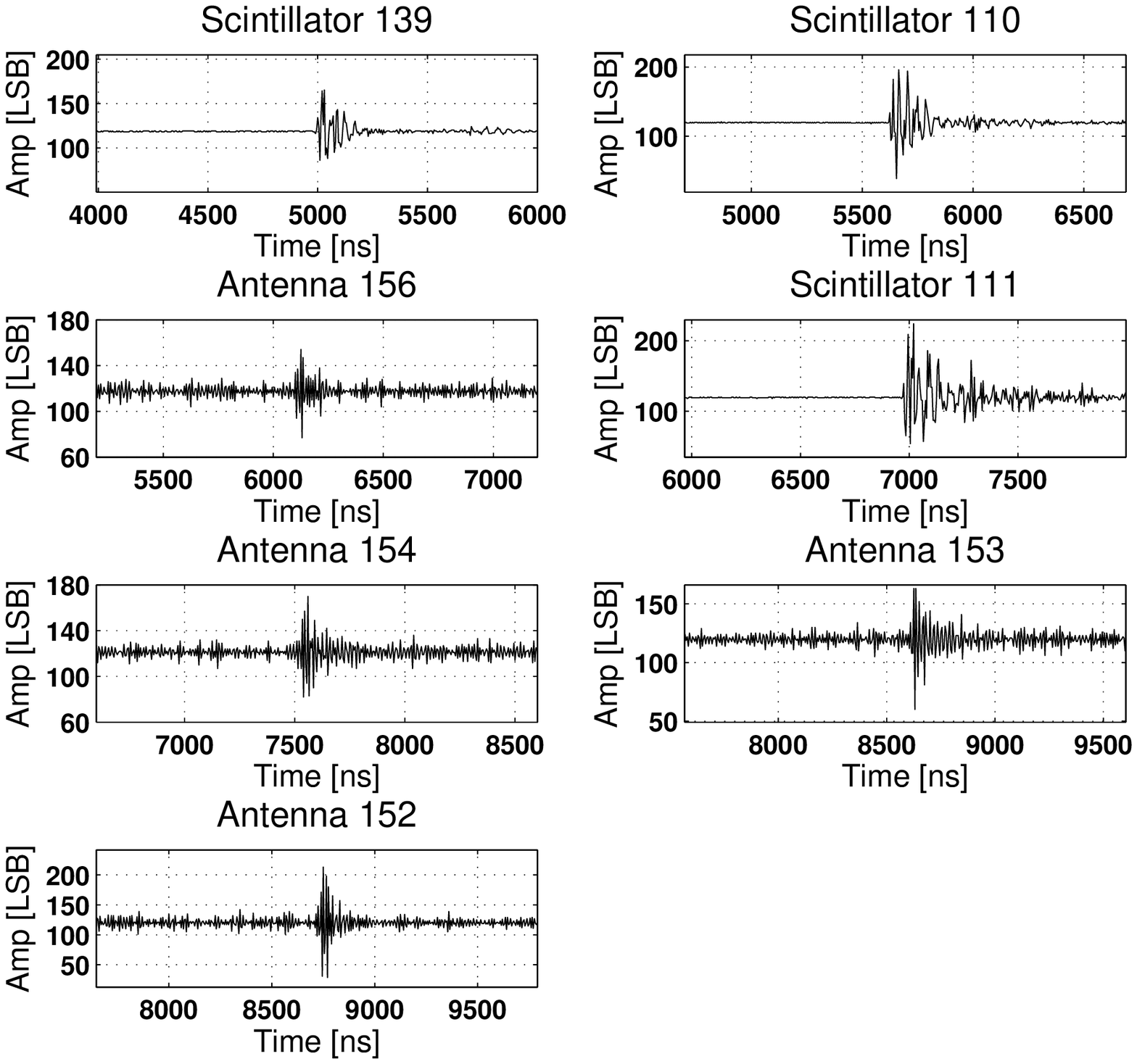}
\end{center}
\caption{ \label{fig:coincidence-signal} \textit{ Waveforms for the hybrid event A, involving 4 antennas and 3 scintillators in coincidence. Delays induced by signal propagation through the fiber and cables are not taken into account for in this display. }}
\end{figure}

In January 2010, 15 TREND antennas were set up around the cross-point of the 21CMA baselines (see \autoref{fig:crosspoint-layout}), covering a total area of 350~m$\times$800~m $\sim$ 0.2~km$^2$. Data have been recorded with this setup since then, and the cosmic ray candidates search procedure described in
\autoref{sec:selection-cuts} has been applied to the reconstructed events. \\
\indent An array composed of three scintillator detectors was installed at the same location. The three detectors are separated by $\sim$200~m from one
another. Each of them is composed of a 0.5~m$\times$0.5~m$\times$2~cm standard plastic scintillator in direct view of a Photonis XP2020 photomultiplier tube (PMT). The PMT signal is converted by an optical transmitter working in the 20-200~MHz frequency range and sent to the acquisition room through an optical fiber. Similarly to an antenna signal, it is then digitized by an 8-bits ADC
working at a rate of 200~MSamples/s. A triggering procedure similar to the one described in \autoref{sec:daq-description} is applied to the scintillators signals. The value of the factor $N$ is set in order to obtain an individual trigger rate around 25~Hz for each scintillator. A selection of triggers in coincidence on the 3 scintillators is performed in an off-line analysis procedure. \\
%
\indent The two systems have run in coincidence for 19 live days so far. During this period, 620 3-fold scintillator events were recorded. The rate of random 3-folds coincidences being lower than 3$\times10^{-3}$/day according to calculations (see \ref{sec:appendix-B}), these coincident events have to correspond to EAS in their vast majority. A reconstruction of the direction of origin was performed for these events a
ssuming a plane wavefront. The distribution of these reconstructed directions is compatible with what is expected for cosmic rays (see \autoref{fig:scints}). \\
%
\indent During this period of hybrid operation, three cosmic ray candidates selected in the radio antenna data following the procedure described in \autoref{sec:selection-cuts} were recorded in coincidence with three scintillators (see \autoref{fig:coincidence-signal}), and two with two scintillators. We have carefully examined these coincidences and the various hypotheses for their origin.

\indent Firstly, we can exclude the occurrence of random coincidences. The rate of random coincidences between two systems $A$ and $B$, triggering on
uncorrelated stationary noises, is computed in \ref{sec:appendix-B}. It writes as:
%
\begin{equation}
\label{eq:two-random-coincidences}
f_{rdm}=2\frac{f_Af_B}{f_A+f_B}(1-\exp(-(f_A+f_B)\Delta{t})),
\end{equation}
with $f_A$ and $f_B$ the trigger frequencies in $A$ and $B$, and $\Delta{t}$ the time window considered for a coincidence. Over the 19 days of data-taking, 620 events were detected in coincidence on the 3 scintillators, corresponding to a trigger rate $f_A$~=~3.8$\times10^{-4}$~Hz. It has also been observed in
practice that $f_B$ = 10~Hz is a safe upper limit for the rate of coincidences on 4 antennas or more. Injecting these values in \autoref{eq:two-random-coincidences}, it is found that the rate of random coincidences between the 3 scintillators and 4 antennas in a time window of 2 $\mu$s can be estimated at less than one per year. The chance to observe within 19 days two random coincidences between two scintillators and four antennas or more is also negligible. The five hybrid coincidences detected by TREND therefore have to be related to the same physical source. \\
%
\begin{table}
\begin{center}
\begin{tabular}{c|cl|cl|cl}
\hline
\hline
  & \multicolumn{2}{c|}{\bf Radio antennas}  &  \multicolumn{2}{c|}{\bf Scintillators}  \\
\hline
\hline
{\bf Event} & $\theta$, & $\phi$    & $\theta$,  & $\phi$       \\
\hline
 A  & 52$\pm$1, & 195$\pm$2 &  49$\pm$3, & 191$\pm$4    \\
 B  & 61$\pm$3, & 359$\pm$2 &  67$\pm$5, &   3$\pm$4    \\
 C  & 42$\pm$1, &  55$\pm$4 &  36$\pm$3, &  56$\pm$5    \\
\hline
\hline
\end{tabular}
\end{center}
\caption{ \label{tab:coincidences-reconstruction} \textit{
  Reconstructed zenith and azimuth angles (in degrees) for the 3 hybrid events detected with the TREND setup (plane wave hypothesis). The direction reconstruction has been performed independently for the antenna and ground array data. All results are statistically compatible. The uncertainties for the antenna and ground array direction reconstructions are estimated using \autoref{eq:standard-theta-phi}, by assuming a time resolution of 10 and 20~ns for the antennas and scintillators respectively. }}
\end{table}
\indent Secondly, the influence of PMT radiations on the antennas can also be excluded. For four out of these five coincidences indeed, antennas
triggered before the scintillators. The time ordering of the triggers is also inconsistent with the transit time of an electromagnetic
wave from the PMTs to the antennas in all five cases. \\
\indent At last, direction reconstructions were performed separately  for the radio and scintillator data for the three events with three scintillators. The results of these two independent reconstructions are in excellent agreement within uncertainties for the three events (see \autoref{tab:coincidences-reconstruction}). This agreement gives us strong confidence that these events have indeed been induced by EAS. In addition, for the two coincidences with two scintillators, the scintillators trigger times are consistent with the values expected from the direction of propagation of the wave reconstructed from the radio data. \\
\indent These five coincident events therefore constitute an unquestionable proof that autonomous radio-detection of cosmic rays has been performed by TREND, and that the selection procedure defined in \autoref{sec:selection-cuts} has allowed to single them out from the background events. Apart from some partial measurements~\cite{Ardouin:2005} quoted earlier, our work establishes for the first time a reliable background rejection method allowing the operation of radio antennas in a stand-alone mode.

\section{Outlook}
\label{sec:outlook}
The ultimate goal of the funded TREND project is the search for signatures of cosmic neutrinos with energies higher than 10$^{16}$-10$^{17}$~eV. We think that the 21CMA site is very well suited for this search. As pointed out in \autoref{sec:status}, it is indeed surrounded by close and high mountain ranges situated $\sim$100~km north of a vast depression. This environment results in rock thicknesses of several tens of kilometers and atmosphere lengths of several kilometers (depending on the observed direction). For energies above 10$^{17}$~eV, this topology favors the observation of $\tau$-induced showers~\cite{Brusova:2007,Ardouin:2009b,Carloganu:2009}. In addition to this and its excellent electromagnetic environment (see \autoref{21cma} and \autoref{sec:antenna-sensitivity}), the TREND site may also benefit from a very low thunderstorm or lightning activity~\cite{thunder}. These features, together with the low cost, easy deployment and maintenance of the TREND detection equipments, could be decisive elements for the success of the TREND project. \\
\indent Several adaptations are however necessary to complete the TREND setup. An antenna design with an optimized sensitivity along the horizon is in particular required, as well as the development of appropriate signal identification and background rejection procedures. These various elements are currently under study and  will be presented in a forthcoming paper~\cite{TREND:2011}, together with an estimation of the TREND sensitivity to UHE neutrinos. The purpose of the present paper is to state the excellent electromagnetic conditions at the TREND site and demonstrate our ability to perform the self-triggering detection of EAS, a prerequisite to the final goal of the Tianshan Radio Experiment for Neutrino Detection project.

\section{Conclusion}
We have given a detailed presentation of the on-going TREND experiment. We described the analysis procedure used for the data recorded in 2009 on a 6-antenna prototype working in a stand-alone mode. Using successive rejection criteria to isolate true physical events, we have selected 25 events which present all characteristics of cosmic-ray induced signals. We have shown the ability of our setup and analysis procedure to perform the autonomous radio-detection of EAS by using a scintillator array running independently in coincidence with an upgraded version of the TREND prototype. With this hybrid setup, several cosmic ray candidates selected from the radio data were independently detected by the scintillator array. Following the  first results obtained on small self-triggering radio-antenna set-ups~\cite{Ardouin:2005, Revenu:2008, Revenu:2010}, TREND has here proposed and validated an identification method for EAS which clearly demonstrates the possible use of a radio-antenna array in a stand-alone mode for their investigation.
Based on these observations and the fulfillment of prerequisites regarding the electromagnetic and topological environments of the TREND experiment site, it is now our goal to exploit this analysis procedure with an extended antenna array dedicated to the identification of UHE neutrinos.

\section*{Acknowledgments}
This work is supported by the France-China Particle Physics Laboratory, the Natural Sciences Foundation of China under Grant No.~10725524, 
No.~10778708 and No.~11050110114, and the Ministry of Science and Technology of China under Grant No.~2009CB824900.

\appendix
\section{Error propagation for the plane wave reconstruction}
\label{sec:appendix-A}

Let us assume that we have a set of $n$ detector units located at positions $\vec{X_i}$, the arrival time of a plane wave at these locations at times $t_i$ with $i=1, ..., n$. Let us further assume that the uncertainties on the unit detectors positions and on the wave propagation speed can be neglected. We write $\Delta t_i = t_i - \hat{t_i}$ the errors on the arrival times, with $\hat{t_i}$ being the true arrival time of the plane wave. Similarly, the angular errors on the plane wave direction (parametrized in spherical coordinates) are written $\Delta \theta = \theta - \hat{\theta}$ and $\Delta \phi = \phi - \hat{\phi}$. At leading order the error $\Delta \vec{k}$ on the wave vector $\vec{k}$ writes:
%
\begin{equation}
\Delta\vec{k} = \partial_{\theta} \vec{k} \, \Delta \theta + \partial_{\phi} \vec{k} \, \Delta \phi + \hat{\vec{k}},
\end{equation}
where $\partial_x$ is the partial derivative operator as $x$. By definition, $\hat{t_i}$ and $\hat{\vec{k}}$ cancel each of the terms of the sum in \autoref{eq:plane-residuals} such that, at leading order, $F_{plane}( t_i; \theta, \phi ) = F_{plane}( \Delta t_i; \Delta \theta, \Delta \phi)$. Furthermore, $\theta$ and $\phi$ are calculated by minimizing the plane wave residuals, $F_{plane}$, hence $\Delta \theta$ and $\Delta \phi$ are related to $\Delta t_i$'s through:
%
\begin{equation}
\label{eq:plane_condition}
\partial_{\theta} F_{plane} = \partial_{\phi} F_{plane} = 0,
\end{equation}
which yields a linear set of equations. Inverting the latter equations for various $t_i$'s one finds the error propagation coefficients, written as:
%
\begin{eqnarray}
\partial_{t_k} \theta & = & \frac{n}{\Delta} \left( a_{k,0,\theta} \, b_{\phi,\phi} - a_{k,0,\phi} \, b_{\theta,\phi} \right) \\
\label{error-phi}
\partial_{t_k} \phi   & = & \frac{n}{\Delta} \left( a_{k,0,\phi} \, b_{\theta,\theta} - a_{k,0,\theta} \, b_{\theta,\phi} \right),
\end{eqnarray}
with
%
\begin{eqnarray}
a_{i,j,x} = ( \vec{X_i} - \vec{X_j} ) \cdot \partial_x \vec{k} \\
\label{eq-b-factor}
b_{x,y} = \sum_{i=1}^{n-1}\sum_{j=i+1}^{n} a_{i,j,x} \, a_{i,j,y}.
\end{eqnarray}
By extension we write $\vec{X_0} = \sum \vec{X_i}/n$, the middle position of the set of detector units. The determinant $\Delta = b_{\theta, \theta} \, b_{\phi, \phi} - b_{\theta, \phi}^2$ falls from the inversion of the equation set~\ref{eq:plane_condition}. \\

\subsection{Standard deviation on angular coordinates}
In the particular case of independently distributed time errors of standard deviation $\sigma_{t_k}$, the resulting standard deviations $\sigma_{\theta}$ and $\sigma_{\phi}$ on $\theta$ and $\phi$ at leading order can be written as a squared sum of the error propagation coefficients~:
%
\begin{eqnarray}
\label{eq:standard-theta-phi}
 \sigma_{\theta}^2 = \sum_{k=1}^{n} \left( \partial_{t_k} \theta \right)^2 \sigma_{t_k}^2 & , &
 \sigma_{\phi}^2   = \sum_{k=1}^{n} \left( \partial_{t_k} \phi \right)^2 \sigma_{t_k}^2.
\end{eqnarray}
Let us further define the absolute pointing accuracy as $\sqrt{\left<\psi^2\right>}$, with $\psi$ the angle between the true wave direction and the reconstructed one. At leading order, $\psi^2$ is linearly related to $\Delta \theta^2$ and $\Delta \phi^2$ and the pointing accuracy simply goes as:
%
\begin{equation}
  \left< \psi^2 \right> = \sigma_{\theta}^2 + \sin^2(\theta) \sigma_{\phi}^2.
\end{equation}

\subsection{Statistical significance of the residuals}
Further expanding $\Delta \theta$ and $\Delta \phi$ at leading order in $\Delta t_i$'s, the plane wave residuals, $F_{plane}$ is:
%
\begin{equation}
F_{plane} = \Delta T^t \, C \, \Delta T,
\end{equation}
where $\Delta T$ is a size $n$ vector of the $\Delta t_i$'s. $C$ is a $n \times n$ correlation like matrix, whose coefficients are given as:
%
\begin{eqnarray}
 c_{k,l} & = &
 \partial_{t_k} \theta \, \partial_{t_l} \theta \, b_{\theta, \theta} +
 \partial_{t_k} \phi \, \partial_{t_l} \phi \,  b_{\phi, \phi} +
 (
 \partial_{t_k} \theta \, \partial_{t_l} \phi + \notag \\
 &  &
 \partial_{t_k} \phi \, \partial_{t_l} \theta
 ) b_{\theta, \phi} -
 n \, (
 a_{k,0,\theta} \, \partial_{t_l} \theta \ + a_{l,0,\theta} \, \partial_{t_k} \theta + \label{eq:plane-residual-matrix} \\
 &  &
 a_{k,0,\phi} \, \partial_{t_l} \phi \ + a_{l,0,\phi} \, \partial_{t_k} \phi \
 ) +
 n \, \delta_{k,l} - 1 \notag,
\end{eqnarray}
where $\delta_{k,l}$ is the Kronecker $\delta$-function. \\
\indent Let us consider the particular case of independent and identically distributed ({\it iid}) time errors, $\Delta t_i$, following a centered Gaussian law of standard deviation $\sigma_t$. The matrix $C$ is symmetrical and positive-defined, hence it diagonalizes over an orthonormal basis, $u_i$, yielding~:
%
\begin{equation}
 F_{plane} = \sum_{i=1}^n \lambda_i u_i^2,
\end{equation}
with $\lambda_i \geq 0$ the eigen values of $C$. Following from orthonormality, the base vectors $u_i$ are also centered Gaussians {\it iid} with the same standard deviation $\sigma_t$. Consequently, the normalized plane wave residuals $F_{plane}/\sigma_t^2$ are distributed as linear combinations of $\chi^2$ variables, with positive coefficients. The cumulative density function, $P(F_{plane} \leq c )$ can efficiently be computed using Ruben's expansion in $\chi^2$ series~\cite{Ruben:1962,Kotz:1967}. See for example the algorithm AS~$204$~\cite{Farebrother:1984} for a numeric implementation. This procedure provides the statistical significance or p-value $P( F_{plane} \leq F_{plane}^{obs} )$ of the measured plane wave residuals, $F_{plane}^{obs}$ when testing the hypothesis \{${\cal H}_0$: {\it the wavefront is plane like with centered Gaussian iid time errors of standard deviation $\sigma_t$}\}.

\section{Random coincidences}
\label{sec:appendix-B}
Let us first consider the case of a single detection unit triggering on a random stationary noise at a rate $f$. Note that the detection unit might be a set of various sub-systems as long as the triggers it provides can be considered as stationary. Starting the observation at an arbitrary time $t_0$, we recall that the elementary probability, $dp$ for the first trigger to occur exactly in the time interval $[t;t+dt]$, $t \geq t_0$, goes as:
%
\begin{equation}
\label{eq:single-dp}
dp = f \, \exp(-f ( t - t_0 ) ) \, dt,
\end{equation}
while the probability, $p$, to have no triggers in the time interval $[t_0;t]$ is given as:
%
\begin{equation}
\label{eq:single-no-P}
p = \exp(-f ( t - t_0 )).
\end{equation}

\subsection{Random coincidence probability}
Let us now consider the case of $n$ detector units triggering on uncorrelated stationary noises at rates $f_i$~($i \in [1,n]$). Starting the observation at time $t_0 = 0$ let us assume that the following event occurred: \{${\cal E}_0$: {\it each receiver triggered only once, at time $t_i$ , and we have $t_0 \leq t_1 \leq t_2 \leq ... \leq t_n$}\}. The event ${\cal E}_0$ is the sum of $n$ independent events \{${\cal E}_i$: {\it the $i^{th}$ trigger occurs at time $t_i$ starting from $t_{i-1}$ and there is no other unit triggering in the time interval $[t_{i-1}; t_i]$}\}. Following from \autoref{eq:single-dp} and \autoref{eq:single-no-P}, the elementary probability, $dp_i$, for the event ${\cal E}_i$ writes as:
%
\begin{eqnarray}
 dp_i & = & f_i \, \exp( -f_i \, ( t_i - t_{i-1} ) ) \, dt_i \,  \prod_{j=1, j \ne i}^n \exp( -f_j \, ( t_i - t_{i-1} ) )  \notag \\
      & = & f_i \, \exp( -f_{\Sigma} \, ( t_i - t_{i-1} ) ) \, dt_i, \label{eq:elementary-dpi}
\end{eqnarray}
where $f_{\Sigma} = \sum_{i=1}^n f_i$. Since the event ${\cal E}_0$ is the sum of the independent events ${\cal E}_i$'s, the elementary probability $d^{n}p$ for the event ${\cal E}_0$ is the product of the elementary probabilities $dp_i$, yielding:
%
\begin{equation}
\label{eq:elementary-dnp}
 d^{n}p = \exp( -f_{\Sigma} \, t_n ) \, \prod_{i=1}^n f_i \, dt_i.
\end{equation}
Integrating the latter probability density over $t_1, t_2, ..., t_{n-1}$ one gets the elementary probability $dp$ for the $n^{th}$ trigger to occur at the time $t_n$, as:
%
\begin{equation}
dp = \int_{0}^{t_2} \int_{0}^{t_3} ... \int_{0}^{t_n} d^{n}p = \left( \prod_{i=1}^n f_i \right) \frac{ t_n^{n-1} }{(n-1)!} \exp( -f_{\Sigma} \, t_n  ) dt_n.
\end{equation}
The probability $P( t_n \leq t )$ that the given succession of triggers has occurred at time $t$ is then given by one more integration of $dp$ over $[0,t]$. It can be checked that this probability does not depend on the time ordering of the $n$ triggers. Hence, the probability is the same for any permutation of the sequence of triggers. It only depends on the time $t_{last}$ of the last trigger and on the rates $f_i$. Therefore, considering the $n!$ permutations of the sequence of triggers the probability $P(t_{last} \leq t)$ that each of the $n$ units triggered {\it once and only once} in the time interval $\Delta t = t - t_0$ is $n! \, P(t_n \leq t)$, yielding:
%
\begin{equation}
\label{eq-p-E0}
P( t_{last}\leq t ) = n! \left( \prod_{i=1}^n p_i \right) \, \left( 1 - T_{n-1}( f_{\Sigma} \, \Delta t ) \exp( -f_{\Sigma} \, \Delta t ) \right).
\end{equation}
where $p_i = f_i / f_{\Sigma}$ is the probability for a trigger to occur on the $i^{th}$ unit and $T_n( x ) = \sum_{k=0}^n x^k/k!$ is the $n^{th}$ order Taylor expansion of the exponential function.

\subsection{Random coincidence rate}
Let us come back to the sequence of $n$ triggers triggering at the times $t_1 \leq t_2 \leq ... \leq t_n$. Taking the $1^{st}$ trigger time, $t_1 = 0$, as the time origin a similar computation as for \autoref{eq:elementary-dpi} and \autoref{eq:elementary-dnp} yields the elementary probability $d^{n}p_1$ for getting the time sequence $t_2 \leq ... \leq t_n$, given that no other trigger occurred on the $1^{st}$ unit. It goes as:
%
\begin{equation}
 d^{n}p_1 = \exp( -f_{\Sigma} \, t_n ) \, \prod_{i=2}^n f_i \, dt_i,
\end{equation}
with $f_{\Sigma}$ the sum of all rates, including the rate $f_1$ for triggering on the $1^{st}$ receiver. Integrating the latter equation one can compute the probability $P_1( t_n \leq t )$ that this trigger sequence has occurred at the time $t$. The rate $R_1( t_n \leq t )$ of such events is then given by the product of the triggering rate $f_1$ on the $1^{st}$ unit times the probability $P_1( t_n \leq t )$ for the right triggering sequence to follow. Once more, one can check that this rate does not depend on the particular ordering of the trigger sequence. Therefore, considering all permutations of the trigger sequence, the rate $R( t_{last} \leq t )$ of coincidences of the $n$ detector units, in a time window $\Delta t = t - t_{first}$, is:
%
\begin{equation}
R( t_{last} \leq t ) = n! \, f_{\Sigma} \left( \prod_{i=1}^n p_i \right) \, \left( 1 - T_{n-2}( f_{\Sigma} \, \Delta t ) \exp( -f_{\Sigma} \Delta t ) \right),
\end{equation}
where it must be understood that it is assumed that each receiver triggered {\it once and only once} in the time window $\Delta t$. Nevertheless,
a similar reasoning as used previously holds that the probability to get one more trigger from any detector unit is negligible within the time window $\Delta t$ provided that the product $f_{\Sigma} \, \Delta t$ is small enough.

\bibliographystyle{h-physrev4}
\bibliography{trend-proto_v2}

\end{document}